\documentclass[english,12pt]{article}
\usepackage[T1]{fontenc}
\usepackage[latin9]{inputenc}
\usepackage[letterpaper]{geometry}
\usepackage{array}
\usepackage{amsmath}
\usepackage{amssymb}
\usepackage{bbm}
\usepackage{subfigure}
\usepackage{graphicx}
\usepackage{epstopdf} 
\usepackage{grffile}
\usepackage{color}
\usepackage{pstricks}
\usepackage{axodraw4j}

\makeatletter

\providecommand{\tabularnewline}{\\}

\usepackage{youngtab}
\usepackage{cite}
\usepackage{a4wide}

\newcommand{\symm}{\tiny\Yvcentermath1\yng(2)}
\newcommand{\fund}{\ensuremath{\Box}}
\newcommand{\afund}{\ensuremath{\overline{\Box}}}

\newcommand{\SU}[1]{\mathrm{SU}(#1)}
\newcommand{\U}[1]{\mathrm{U}(#1)}
\newcommand{\Sp}[1]{\mathrm{Sp}(#1)}

\newcommand{\kappair}{\xi}

\newcommand{\sunc}{\mbox{SU$(\nc )$}}
\newcommand{\sun}{\mbox{SU$(\nnc )$}}

\newcommand{\Uo}{\mathrm{U(1)}}

\newcommand{\wir}{W_{\mbox{\tiny IR}}}
\newcommand{\lir}{\Lambda_{\mbox{\tiny IR}}}
\newcommand{\nir}{n_{\mbox{\tiny IR}}}
\newcommand{\wmag}{W_{\mbox{\tiny mag}}}
\newcommand{\welec}{W_{\mbox{\tiny elec}}}

\newcommand{\wuv}{W_{\mbox{\tiny UV}}}
\newcommand{\thooft}{\mbox{'t~Hooft }}
\newcommand{\luv}{\Lambda_{\mbox{\tiny UV}}}
\newcommand{\nuv}{n_{\mbox{\tiny UV}}}

\newcommand{\zuv}{z_{\mbox{\tiny UV}}}
\newcommand{\ruv}{r_{\mbox{\tiny UV}}}
\newcommand{\auv}{a_{\mbox{\tiny UV}}}
\newcommand{\rir}{r_{\mbox{\tiny IR}}}
\newcommand{\air}{a_{\mbox{\tiny IR}}}
\newcommand{\puv}{\p_{\mbox{\tiny UV}}}
\newcommand{\zzuv}{{\mbox{\small ($z_{\mbox{\tiny UV}}$)}}}
\newcommand{\zir}{z_{\mbox{\tiny IR}}}

\newcommand{\pzuv}{{\mbox{\small ($ipz_{\mbox{\tiny UV}}$)}}}
\newcommand{\pzir}{{\mbox{\small ($ipz_{\mbox{\tiny IR}}$)}}}
\newcommand{\rpzuv}{{\mbox{\small ($pz_{\mbox{\tiny UV}}$)}}}
\newcommand{\rpzir}{{\mbox{\small ($pz_{\mbox{\tiny IR}}$)}}}
\newcommand{\bcft}{b_{\mbox{\tiny CFT}}
}

\newcommand{\sunf}{\mbox{SU$(\nf )$}}
\newcommand{\nc}{{N}}
\newcommand{\nnc}{{n}}
\newcommand{\mc}{{M}}
\newcommand{\mmc}{{m}}
\newcommand{\nf}{{{F}}}

\newcommand{\pp}{\varphi}

\newcommand{\p}{\Phi}

\newcommand{\beq}{\begin{equation}}
\newcommand{\eeq}{\end{equation}}
\newcommand{\ba}{\begin{eqnarray}}
\newcommand{\ea}{\end{eqnarray}}
\newcommand{\nn}{\nonumber}

\newcommand{\tr}{\operatorname{tr}}

\makeatother

\usepackage{babel}

\begin{document}


\begin{titlepage}

{\flushleft{\small IPPP/10/86 ; DCPT/10/172  \\ CERN-PH-TH/2010-234}}\\[0.5cm]

\begin{center}

 {\LARGE\bf A slice of AdS$_5$ as the large $\nc$ limit\\
 \vspace{0.3cm}  of Seiberg duality}\\[0.8cm]



{Steven Abel$^{a,b,}$\footnote{s.a.abel@durham.ac.uk} and Tony Gherghetta$^{c,}
$\footnote{tgher@unimelb.edu.au}\\
\vspace{4ex}
\emph{$^a$\small Institute for Particle Physics Phenomenology}\\
\emph{\small Durham University, DH1 3LE, UK}\\
\vspace{1ex}
\emph{$^b$\small Theory Division, CERN, CH-1211 Geneva 23, Switzerland}\\
\vspace{1ex}
\emph{$^c$\small School of Physics, University of Melbourne}\\
\emph{\small Victoria 3010, Australia}}

 
\begin{abstract}
\vspace{-0.3cm}
\noindent 
A slice of AdS$_5$ is used to provide a 5D gravitational description of 4D 
strongly-coupled Seiberg dual gauge theories. An 
(electric) SU$(\nc )$ gauge theory in the conformal window at large $\nc$ is 
described by the 5D bulk, while its weakly coupled (magnetic) dual is confined to the IR brane. 
This framework can be used to construct an ${\cal N} = 1$ MSSM on the IR brane, 
reminiscent of the original Randall-Sundrum model. In addition, we use our 
framework to study strongly-coupled scenarios of supersymmetry breaking 
mediated by gauge forces. This leads to a unified scenario that connects the 
extra-ordinary gauge mediation limit to the gaugino mediation limit in warped space.
\end{abstract}

\end{center}
 
\end{titlepage}
\setcounter{footnote}{0}

\section{Introduction}

\vspace{-0.3cm}

Seiberg duality \cite{S:Duality,Intriligator:1995au} is a powerful tool for studying strong dynamics, enabling calculable 
approaches to otherwise intractable questions, as for example in the discovery of metastable minima in the free 
magnetic phase of supersymmetric quantum chromodynamics (SQCD)~\cite{ISS} (ISS).
Despite such successes, the phenomenological applications of Seiberg duality are somewhat limited, simply by the 
relatively small number of known examples. 

An arguably more flexible tool for thinking about strong coupling is the gauge/gravity correspondence,
namely the existence of gravitational duals for strongly-coupled gauge theories~\cite{Maldacena:1997re,Gubser:1998bc,Witten:1998qj}. Such duals are known in certain cases to be equivalent to a cascade of 
Seiberg dualities, the prototypical example being an $\SU{N+M}\times\SU{N}$ 
theory~\cite{Klebanov:2000nc,Klebanov:2000hb}, which describes the moduli space of a theory of branes and $M$ 
fractional branes on the warped-deformed conifold (see \cite{Aharony:1999ti,Strassler:2005qs} for reviews). 
Warped geometries are indeed generally characteristic of compactifications with flux \cite{Giddings:2001yu}. 
Metastable ISS-like supersymmetry breaking can be 
implemented by having various different brane configurations at the 
end of such throat-like geometries, as in for example~\cite{Kachru:2002gs,Franco:2005zu,Franco:2006es,Argurio:2006ny}, and can be successfully mediated as in \cite{GarciaEtxebarria:2006rw,Benini:2009ff,McGuirk:2009am}.

It is natural to suppose that these and similar examples can be approximated by the ``slice-of-AdS'' 
Randall-Sundrum scenario~\cite{Randall:1999ee} (RS1). The RS1 formulation provides an ideal framework
for discussing the effects of strong coupling without having to use the full complexity of string theory. 
Indeed the phenomenology is typically dominated by the low energy modes, and the throat 
itself enters mainly through the Kaluza-Klein (KK) spectrum. Moreover the latter tends to be linear, 
so that certain phenomena can be rather universal.
RS1 can therefore be a useful approximation because precise knowledge of the gravitational dual is often unnecessary, 
aside from its general scale and warping. 
One example, again in the context of supersymmetry breaking, is gaugino 
mediation~\cite{Mirabelli:1997aj,Kaplan:1999ac,Chacko:1999mi,Gherghetta:2000kr,Csaki:2001em,Chacko:2004mi,McGarrie:2010kh,Green:2010ww,McGarrie:2010qr,Sudano:2010vt}. 
There supersymmetry breaking is mediated to scalars localized on the UV brane via bulk gauge modes. However, the resulting suppression of the scalar masses is relatively independent of whether the bulk is flat or warped, and depends only on the separation in the approximately linear KK spectrum. 

On the other hand there is an obvious drawback of RS1: it is unclear whether the strongly coupled 4D system 
that is supposed to correspond to a given configuration actually exists. Because of this it seems interesting to study 
strongly coupled 4D field theories that 
have an RS1-like configuration, namely a period of conformal running with a well behaved and calculable 
perturbative theory in the infra-red. In particular, we would like to {\em begin} with 
simple ${\cal N}=1$ 4D dynamics that has certain properties (such as supersymmetry breaking), 
and by taking a strong coupling limit introduce desirable features of extra dimensional physics (such as gaugino mediation).

\newpage

This paper presents a straightforward method based on  Seiberg duality by which such 
models can be constructed. It works as follows. Suppose one is interested in reproducing an RS1 configuration 
that has particular weakly coupled theories located on the IR brane, and the UV brane. The holographic principle 
tells us that the UV theory should exist as elementary degrees of freedom in the 4D theory whereas the IR theory should be composite. Hence the IR theory (including any gauge degrees of freedom) can be identified as the 
free-magnetic dual of an electric/magnetic pair. Assuming for definiteness that the theory is ${\cal N}=1$, $\sun$ 
and vector-like, then there should be $\nf > 3 \nnc$ fundamental flavours in the IR theory for it to be IR-free.
This theory becomes strongly coupled at some scale $\lir$, above which an 
electric theory takes over. Many different (indeed an infinity of) electric theories
flow to this particular magnetic theory. The canonical one is an asymptotically free
${\cal N}=1$ $\sunc$ theory, where $\nc=\nf-\nnc$. However, consider instead an $\sunc$ electric theory 
 with $\nf_1=\Delta_F+\nf$ flavours, of which $\Delta_F$ have 
mass $\lir$, with the rest being massless. If we choose $\frac{3}{2}\nc < \nf_1<3\nc $, then  this theory 
is in the conformal window. Above the mass-scale  $\lir$ the theory flows to a fixed point and can enjoy an 
arbitrarily long period of conformal running. At the scale  $\lir$ however, we integrate out the $\Delta_F$ heavy flavours, and the theory flows to the same weakly coupled magnetic theory in the IR. Moreover the $\sunc$ theory approaches strong coupling as $\nf_1 \rightarrow \frac{3}{2}\nc $. By taking a large $\nc$ limit and adding massive and massless flavours to keep the ratio $\nc/\nf_1$ fixed (the Veneziano limit~\cite{Veneziano:1976wm}), one can arrange for the conformal fixed point to be at arbitrarily strong coupling.  In this limit, we replace the period of conformal running with a bulk gravitational dual, and arrive at an RS1 configuration, with the global flavour symmetries becoming bulk gauge symmetries, the magnetic theory confined to the IR brane, and any 
elementary degrees of freedom to which the theory weakly couples, becoming degrees of freedom on the 
UV brane. In a sense this is simply the shortest possible cascade, in which strong and weak coupling are connected by a single Seiberg duality.

Our bulk is required to be a slice of the gravitational dual of strongly coupled SQCD, whatever that may be. 
Since the ${\cal N}=1$ conformal theory has an $R$-symmetry, a good candidate for the geometry is six 
dimensional, namely $AdS_5 \times S_1$ with the $R$-symmetry corresponding to translations along the circle,
$S_1$. A solution of this type for general $\nc,\nf$ was constructed in Ref.~\cite{Klebanov:2004ya} using the 
effective action of six-dimensional non-critical superstrings. This specific background can be realized 
with a stack of D3 branes at the tip of the SL$(2,R)/\Uo$ cigar, with the flavours of fundamental and antifundamental being provided by space-time filling uncharged D5 branes.

It should be noted that this background is not under good theoretical control. Its curvature is 
large so the solution will get 
order one corrections from higher-derivative terms in the effective action. Indeed 
as argued in Ref.~\cite{Bigazzi:2005md} and also in Section~\ref{gauging}, 
it is likely that there can {\em never} be a weakly curved gravitational dual 
of SQCD in the conformal window.
Nevertheless, it is also likely that the effect of the large curvature will be to change the 
parameters of the solution in Ref.~\cite{Klebanov:2004ya} while leaving its general properties intact~\cite{Polyakov:1998ju,Polyakov:2000fk}. As we have said, for phenomenological purposes these general properties can often be sufficient.

We will present in this paper two applications that illustrate the approach. Following a summary of the 
Renormalisation Group (RG) properties of SQCD in the next section, and more details of the configuration in 
Section~\ref{5ddual}, 
we will construct an ${\cal N}=1$ MSSM-on-the-IR-brane type of 
model, reminiscent of the original RS1 configuration. We begin with the basic MSSM and find the electric theory by 
identifying $\SU{2}_L$ with $\Sp{1}$, and then performing an $\Sp{\nc}$ type Seiberg duality. In order to be able to put the 
electric theory in the conformal window and then take the large $\nc$ limit, the only modification we need to make to the 
spectrum of the usual MSSM is to include an arbitrary number of 
heavy Higgs pairs. The left-handed states are all identified as composite states, as are the gauge degrees of freedom of
$\SU{2}_L$. The right-handed fields on the other hand can be an arbitrary mixture of elementary and composite states
(with the latter being identified as the ``mesons'' of the Seiberg duality). The remaining $\SU{3}_c\times \U1_Y$ gauge degrees of freedom are bulk modes. 

The second application is to gauge mediation of supersymmetry breaking. There has been recent interest 
in 4D models that can reproduce the phenomenology associated with gaugino mediation, namely 
gaugino masses that are much heavier than scalar masses at the mediation scale~\cite{Mirabelli:1997aj,Kaplan:1999ac,Chacko:1999mi,Gherghetta:2000kr,Csaki:2001em,Chacko:2004mi}. While this set-up is 
simple to configure in extra-dimensional models, the underlying supersymmetry breaking has to be added by hand. 
The recent interest has therefore been in finding models in which the dynamics of supersymmetry breaking is 
included~\cite{McGarrie:2010kh,Green:2010ww,McGarrie:2010qr,Sudano:2010vt}. 
Here we begin with the ``simplified gauge mediation''
scenario of Murayama and Nomura~\cite{MN}. We show that taking the large $\nc$ limit of this theory as described above 
yields a gaugino mediation model in AdS, with metastable 
supersymmetry breaking on the IR-brane, matter and messenger fields on the 
UV brane and gauge fields in the bulk, a scenario that has been considered in a number of phenomenological 
applications~\cite{Dudas:2007hq,Abel:2010uw}. As in conventional gaugino mediation, the scalar mass squareds are generated by (5D) one loop diagrams that are relatively suppressed due to the bulk separation of matter fields and supersymmetry breaking. We find that (in a certain limit) the suppression is of the form $m_i^2 \sim M_\lambda^2 /\bcft$ where 
$\bcft$ is the bulk contribution to the 
beta function coefficient, which effectively counts the messenger content of 
the strongly coupled CFT.
This is the strong coupling version of the usual $m_i^2 \sim M_\lambda^2 /N_{\mbox{\tiny mess}}$ relation that one finds in perturbative extra-ordinary gauge mediation~\cite{Cheung:2007es}. From this ``extra-ordinary gauge mediation limit'', we will show how one can go continuously to the opposite extreme of supersymmetry breaking in AdS by twisted boundary conditions~\cite{Gherghetta:2000kr}, which we refer to as the ``gaugino mediation limit''.

\section{UV completions of weakly coupled SQCD}

\label{UVcompletion}

We begin by discussing the possible UV completions of SQCD from the 
Seiberg duality point of view. It is useful to study the flow down from high energies. 
In the far UV we assume a standard ${\cal N}=1$ 
supersymmetric QCD  theory which we refer to as the ``electric theory''. 
This is an $\sunc$
theory with $\nf$ flavours \cite{S:Duality,Intriligator:1995au}. 
With no superpotential this theory
has a global $\sunf_{L}\times\sunf_{R}\times\Uo_{B}\times\Uo_{R}$
symmetry. These global symmetries are anomaly free with respect to
the gauge symmetry. There is also an anomalous $\Uo_{A}$ symmetry
that will be irrelevant for our discussion. The particle content is shown in Table~\ref{sqcd0}.
\begin{table}
\centering{}\begin{tabular}{|c||c|c|c|c|c|}
\hline 
$ $ &  $\SU\nc$ & $\sunf_L$ & $\sunf_R$ & $\U1_{B}$ & $\U1_{R}$\tabularnewline
\hline
\hline 
$Q$ & $\fund$ & $\fund$ & 1 & $\frac{1}{\nc}$ & $1-\frac{\nc}{\nf}$\tabularnewline
\hline 
$\tilde{Q}$ &$\afund $& 1 & $\afund$ & $-\frac{1}{\nc}$ & $1-\frac{\nc}{\nf}$\tabularnewline
\hline
\end{tabular}\caption{\emph{Spectrum and anomaly free charges in }SQCD\emph{.}\label{sqcd0}}
\end{table}

Although the features of the RG flow of these theories will be well known to many readers, it is for clarity
worth recapping those elements that we need for our discussion\footnote{The behaviour of this theory under RG flow has been described in many excellent texts, for example Ref.\cite{Strassler:2005qs}.}.
The coupling runs according to the exact NSVZ $\beta$-function~\cite{NSVZ} 
\beq
\label{nsvz}
\beta_{\frac{8\pi^2}{g^2}} = \frac{3\nc - \nf(1-\gamma_Q)}{1-\frac{\nc g^2 }{8\pi^2}   }\eeq
where $\gamma_Q$ is the anomalous dimension of the quarks, given at leading order by 
\beq
\label{gamma0}
\gamma_Q = -\frac{g^2 }{8\pi^2}\left( \frac{\nc^2-1}{\nc}\right)\, .
\eeq
Assume that the high energy theory begins in the so-called conformal window,
\beq
\frac{3}{2}\nc < \nf < 3\nc \, .
\eeq
The theory then runs to a fixed point as can be most easily seen when it is 
just inside the conformal window, $\nf = 3(\nc -\nu) $
in the limit where $1\leq \nu \ll \nc$. Solving for the vanishing of the beta function gives
the anomalous dimension of the quarks to be order $\nu$ and a
fixed point at 
\beq
g^{2}_*=\frac{8\pi^2 }{\nc^2-1} \nu\, .
\eeq
Decreasing the number of flavours increases the value of the coupling 
at the fixed point, until in the opposite limit, where $\nf = \frac{3}{2}(\nc +\nu) $,
it approaches strong coupling: solving for the vanishing of the beta function there, one finds that 
\beq
\label{gq-fixed}
\gamma_Q=-1+2\frac{\nu}{\nc}\, ,
\eeq
and therefore by Eq.\eqref{gamma0} we have $ g^{2}_* \nc \gtrsim 8\pi^2$. \footnote{Although 
the anomalous dimension in Eq.\eqref{gamma0} is accurate to one loop, the NSVZ beta function 
in Eq.\eqref{nsvz} is exact. Hence even though the precise form of the anomalous dimension $\gamma_Q$ is unknown, 
as long as it can approach arbitrarily close to $-1$, a fixed point will be found.} 
This result can also be derived from the fact that at a conformal fixed point the $R$-charges and dimensions of 
operators ${\cal O}$ are related as dim~${\cal O}=\frac{3}{2}R_{\cal O}$: this relation and 
the definition
dim~${Q}=1+\frac{\gamma_Q}{2}$ give 
\beq
1+\frac{\gamma_Q}{2}=\frac{3}{2}\left(1-\frac{\nc}{\nf}\right) \, ,
\eeq
the same result as that deduced from the vanishing of the NSVZ beta function in Eq.\eqref{nsvz} 

The theory at the IR fixed point has an equivalent magnetic description.
The magnetic dual theory (which we will call $\overline{\mbox{SQCD}}$)
has a gauged $\SU\nnc$ symmetry, where $\nnc=\nf-\nc$ \cite{S:Duality,Intriligator:1995au}.
Its spectrum is given in Table \ref{sqcd0-1}. (Throughout we will denote
magnetic superfields with small letters and electric superfields with
capitals.)
\begin{table}
\centering{}\begin{tabular}{|c||c|c|c|c|c|}
\hline 
$ $ &  $\SU\nnc$ & $\sunf_L$ & $\sunf_R$ & $\U1_{B}$ & $\U1_{R}$\tabularnewline
\hline
\hline 
$q$ & $\fund$ & $\afund$ & 1 & $\frac{1}{\nnc}$ & $1-\frac{\nnc}{\nf}$\tabularnewline
\hline 
$\tilde{q}$ & $\afund$& 1 & $\fund$ & $-\frac{1}{\nnc}$ & $1-\frac{\nnc}{\nf}$\tabularnewline
\hline 
$\pp$ &1& $\fund$ & $\afund$ & 0 & $2\frac{\nnc}{\nf}$\tabularnewline
\hline
\end{tabular}\caption{\emph{Spectrum and anomaly free charges in} $\overline{\mbox{SQCD}}$\label{sqcd0-1}}
\end{table}
The two theories satisfy all the usual tests of anomaly and baryon
matching if one adds a superpotential \begin{equation}
\label{wmag0}
\wmag=h\, q\pp\tilde{q}.\end{equation}
The equation of motion of the elementary meson, $\varphi$ then projects the superfluous
composite meson $q\tilde{q}$ out of the moduli space of the magnetic
theory. Obviously the magnetic theory is also inside the conformal window, but
where the electric description is strongly coupled the magnetic one is weakly coupled, 
and vice-versa. 

\begin{figure}[h]
\begin{centering}
  \begin{picture}(400,100) (36,0)
    \SetWidth{2.7}
    \Text(7,90)[lb]{(a)}
    \Text(161,90)[lb]{(b)}
    \Text(317,90)[lb]{(c)}
\includegraphics[angle=0,scale=.5]{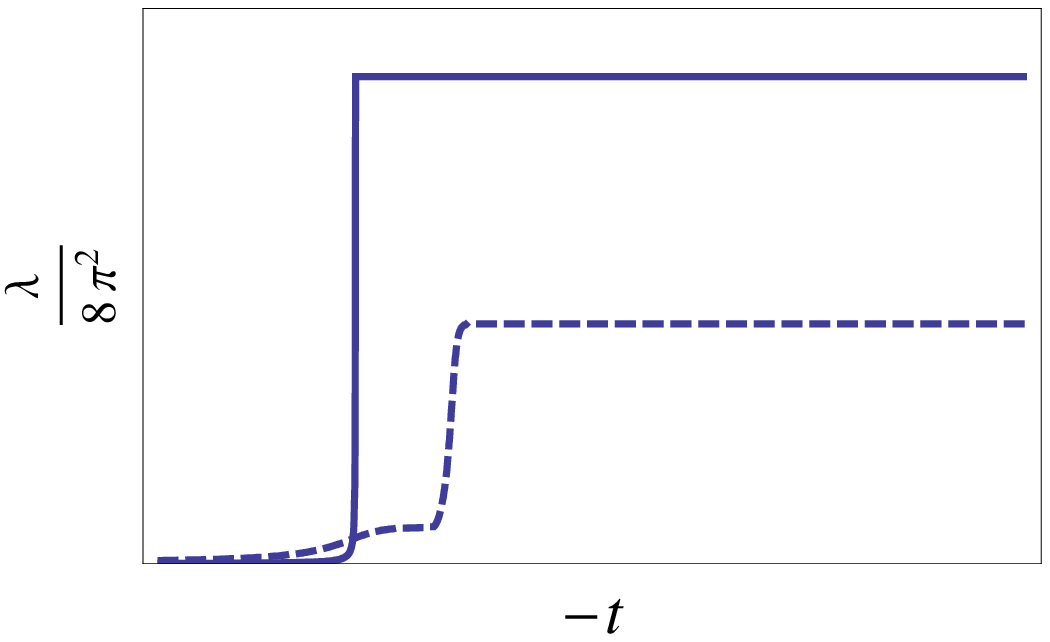}
\includegraphics[angle=0,scale=.5]{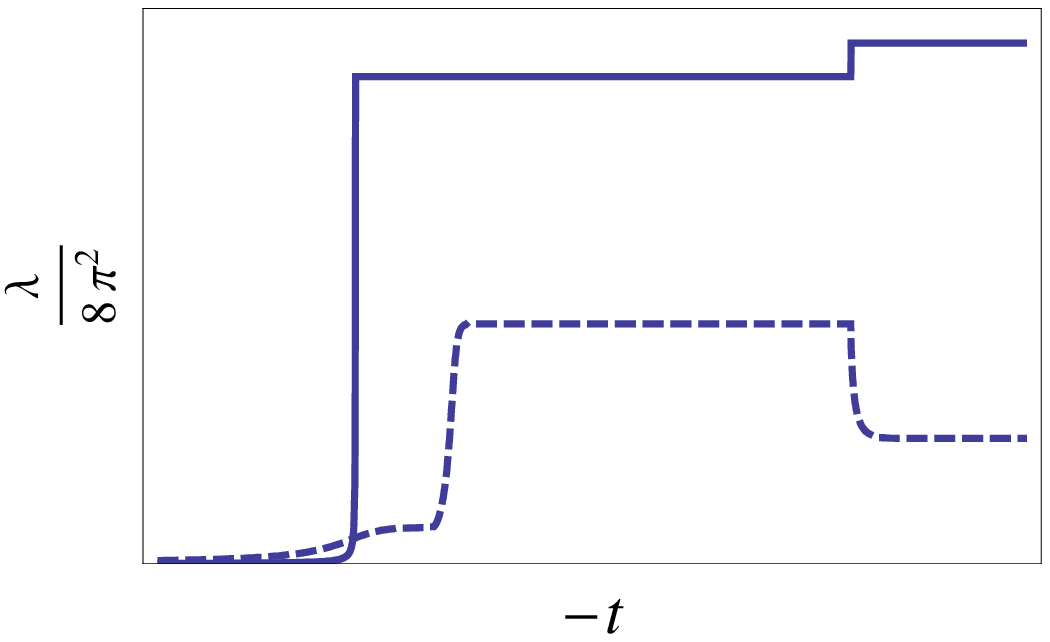}
\includegraphics[angle=0,scale=.5]{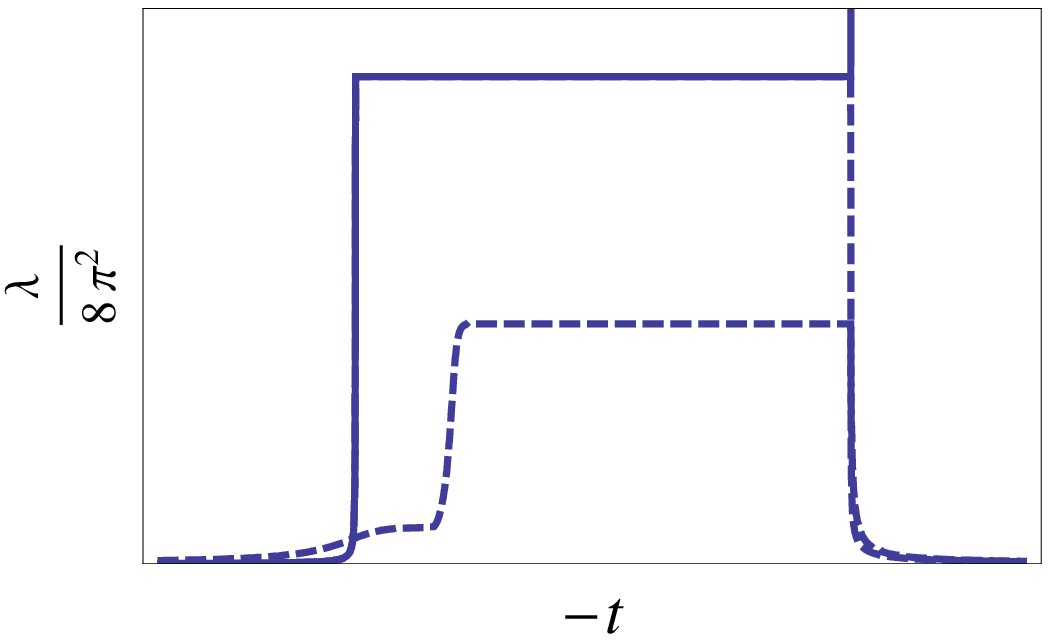}
\end{picture}
\par\end{centering}
\caption{\it Types of SQCD RG flow with $t=\log E$.  The solid and dashed lines are the electric and 
magnetic theories and $\lambda=g^2 \nc$ or $\bar{g}^2 \nnc$ respectively. The undeformed theories flow 
to their conformal fixed points in (a). Upon adding a mass deformation the theories 
flow to new fixed points as in (b) or to IR-free theories as in (c). The magnetic theory was started at small 
coupling in order to show its evolution towards the unstable fixed point first $h_{* }=0$, $g_{* }$
before ending up at the stable fixed point $h_*^{\prime }$, $g_*^{\prime }$.} 
\label{fig:flows}
\end{figure}

Consider the theory with $\nf = \frac{3}{2}(\nc +\nu) $ with $\nc\gg \nu$.
In the absence of the coupling $h$ the $\overline{\mbox{SQCD}}$
theory would clearly be no different from the original SQCD one;
the magnetic 
dual has $\nf = {3}(\nnc -\nu) $ and hence a fixed point at 
\ba
h_*&=&0\nn\\
 \bar{g}^{2}_* &=& \frac{8\pi^2 }{\nnc^2-1} \nu\, .
\ea
This is {\em not} however the fixed point corresponding to that of the electric theory 
and indeed any non-zero coupling $h$ precipitates flow to a new fixed point. The 
dim~${\cal O}=\frac{3}{2}R_{\cal O}$ argument predicts it to be where  
\beq
\gamma_q=1-\frac{3\nnc}{\nf}\,\,\, ; \, \, \gamma_\pp=-2+\frac{6\nnc}{\nf}\, .
\eeq
Note that the beta function of the coupling $h$ is $\beta_h=\frac{h}{2}(\gamma_\pp+2\gamma_q)$ which indeed vanishes 
at this point. The anomalous dimensions themselves are 
\ba
\gamma_q&=&\frac{1}{16\pi^2} \left( \nf h^2 - 2\bar{g}^2 \left( \frac{\nnc ^2 -1}{\nnc} \right) \right) = 1-\frac{3\nnc}{\nf}\nonumber \\
\gamma_\pp &=&\frac{1}{16\pi^2} \left( \nnc h^2 \right) = -2+\frac{6\nnc}{\nf}\, ,
\ea
giving 
\ba
\frac{\nnc (h_*^{\prime })^2}{8\pi^2} &=& \frac{12\nu}{\nf}\nn\\ 
(\bar{g}_*^{\prime })^2 &=& \frac{\nnc(\nnc+2\nf )}{4(\nnc^2-1)} (h_*^{\prime })^2\, .
\ea
When $\nf = {3}(\nnc -\nu) $ and $\nnc\gg \nu$ {\em both} fixed points in the magnetic description are at weak coupling. 
And in the opposite limit where $\nf = \frac{3}{2}(\nnc +\nu) $, {\em both} fixed points are at strong coupling with $\nnc h^2_*\gtrsim 8 \pi^2$.
A numerically evolved example of such flow is shown in Figure~\ref{fig:flows}a: the solid and dashed lines are the flows in the electric and 
magnetic theories described above. The magnetic theory is indeed seen to flow initially towards the unstable fixed point with $h_{* }=0$, $g_{* }$,
before ending up at the stable fixed point $h_*^{\prime }$, $g_*^{\prime }$. The fixed point values of the couplings
indeed obey the above relations.

Now let us consider what happens when we add a relevant deformation to the electric model -- i.e. a new 
term in the superpotential. If this new term breaks the remaining $R$-symmetry the theory will flow, either to a 
new fixed point, or to an IR-free or asymptotically  free theory. The simplest example of such a deformation is a 
quark mass term in the electric theory: 
\beq
\welec=(m_Q)_{i}^j Q^i\tilde{Q}_j\, ,
\eeq
where $i,j$ are flavour indices. Consider the case where $m_Q$ is diagonal of the form 
\beq
m_Q=m_0 \left( 
\begin{array}{ll} 
{\bf 1}_{\nf_1\times\nf_1} & {\bf 0}_{\nf_1\times \nf_2} \nn\\
 {\bf 0}_{\nf_2\times\nf_1} & {\bf 0}_{\nf_2\times\nf_2} 
 \end{array}\right)\, ,
 \eeq
 where $\nf_1+\nf_2=\nf$, and where this value is set at some UV scale, $\luv$.
This term then explicitly breaks both the $R$-symmetry and the global symmetry down to 
\beq
\sunf_{L}\times\sunf_{R}\times\Uo_{B}\times\Uo_{R}\rightarrow \SU{\nf_1}_D\times \SU{\nf_2}_{L}\times\SU{\nf_2}_{R}\times \Uo_{B}\, .
\eeq
At low scales one can integrate out $\nf_1 $ quark/antiquark pairs, 
leaving $\nf_2$ light flavours. The global symmetry is further broken, but a (different) $R$-symmetry is 
 recovered:
\beq
\SU{\nf_1}_D\times \SU{\nf_2}_{L}\times\SU{\nf_2}_{R}\times \Uo_{B}\rightarrow 
\SU{\nf_2}_{L}\times\SU{\nf_2}_{R}\times \Uo_{B}\times \Uo_{R'}\, .
\eeq
Thus provided that $\nf_2>\frac{3\nc}{2}$ the theory remains in the conformal 
 window and flows to a  new fixed point at stronger coupling than the previous one,
 corresponding to $\nf_2$ flavours and $\nc$ colours; this is the solid line in Figure~\ref{fig:flows}b. 
The scale at which one can integrate out the heavy quarks is not simply $m_0$ because the 
quarks have a dimension $\frac{3}{2}(1-\nc/\nf)$ which is different from unity. One finds that the 
mass of the canonically normalised quarks is larger than the energy scale, when the 
latter drops below a value $\lir$ given by 
\beq
\label{eq:lamir}
\lir = m_0 \left( \frac{\lir}{\luv}\right)^{2-\Delta_\pp}\, ,
\eeq
where $\Delta_\pp=\frac{3}{2}(1-\nc/\nf)$ is the dimension of $Q\tilde{Q}$. (Note that at weak coupling when 
$\Delta_\pp = 2+\gamma_\pp$ this gives the usual perturbative approximation 
$\lir = m_0 (1-\gamma_\pp \log(\lir/\luv) )$.) 

\vspace{1cm}

 The corresponding deformation in the magnetic theory is a linear meson term that induces a Higgsing; the magnetic 
 superpotential becomes 
\begin{equation}
\label{wmag}
\wmag=h\, q\pp\tilde{q} -
h \mu_\pp^2 \pp \, ,\end{equation}
where the linear $\pp$ term corresponds 
to the quark mass term and has the same $R$-charge and flavour structure:
\beq
\mu^2_\varphi=\mu_0^2 \left( 
\begin{array}{ll} 
{\bf 1}_{\nf_1\times\nf_1} & {\bf 0}_{\nf_1\times \nf_2} \nn\\
 {\bf 0}_{\nf_2\times\nf_1} & {\bf 0}_{\nf_2\times\nf_2} 
  \end{array}\right)\, .
 \eeq
 This deformation causes $\nf_1$ of the quarks to acquire VEVs 
  of the form 
 \beq
 \langle q \tilde{q} \rangle =\mu^2_\varphi\, .
 \eeq
The gauge symmetry is Higgsed down to $\SU{\nnc_2}$ where $\nnc_2=\nnc-\nf_1= \nf_2-\nc$, and $\nf_2$ 
quarks remain light. (Note that we assume $\nf_1 \leq \nnc$.) 
One can arrange for the change in the number of degrees of freedom to be at the same 
energy scale in both theories by appropriately choosing $\mu_0$. This behaviour is shown in Figure~\ref{fig:flows}b. 

If one chooses $\nf_1$ such that $\nc+1<\nf_2<\frac{3\nc}{2}$ then the behaviour is different again (c.f. Figure~\ref{fig:flows}c). 
At scales below $\lir$ the theory falls out of the conformal 
 window (so to speak) and into the free magnetic range. The gauge coupling of the electric theory hits a Landau pole 
 below $\lir$ and can be matched onto an IR free magnetic description. On the other hand the 
 original magnetic theory flows smoothly (without encountering strong coupling) to the {same} IR free theory. 
 This is classic quasi-conformal (walking) behaviour (see Ref.~\cite{Dietrich:2010yw} for a recent summary).

Note that upon adding the mass deformation the  
electric and magnetic theories describe the same physics at scales much higher than $\lir$ and also in the far IR well below it.
Around $\lir$ however we see that they are very different.
 Thus it is important to realise that these are two physically distinct types of UV completion of a single IR free theory. The first
consists of a Landau pole followed by a relatively strongly coupled conformal UV phase. The second consists of a smooth
 transition to a weakly coupled conformal theory in the UV. A third possible type of UV completion would of course 
 be the asymptotically free electric Seiberg dual of the IR free theory. 
 
\section{A 5D dual description in the strong coupling limit}

\vspace{-0.3cm}

 \label{5ddual}
 
\subsection{The general set-up}
 
 \vspace{-0.3cm}

One of the three possibilities for the UV completion of the deformed SQCD (with a quark mass term
in the electric theory) is a relatively strongly coupled conformal phase in the limit of large $\nc$. 
By the AdS/CFT correspondence~\cite{Maldacena:1997re}, this phase admits a weakly-coupled 
5D gravitational description. Furthermore given that the 4D deformed SQCD is only conformal 
below a UV scale, with conformal invariance broken by a mass deformation in the IR, the dual 
description must be a slice of AdS$_5$. A simple way to mimic these features is to introduce a 
UV brane and an IR brane corresponding to an RS1 scenario~\cite{Randall:1999ee}. This is a 
crude approximation to the underlying dynamics of the microscopic theory. Nevertheless as stressed in the Introduction, many of the qualitative features can be understood in this simplified framework, just as 
AdS/QCD models are thought to encapsulate certain features of QCD. In this subsection 
we wish first to sketch out the general features that such a description must have, based on known 
attempts in the literature to construct gravitational duals of strongly coupled SQCD. 
We then go on to consider the possible 5D configurations that correspond to strongly 
coupled 4D Seiberg duals. 

First the general set-up. We take a slice of AdS and place a UV brane at $\luv$ and an IR brane 
at a confinement scale loosely associated with $\lir$ (we will discuss this point in more 
detail below). In order to accommodate the intrinsic U(1)$_R$ symmetry, the dual 5D description 
must actually be AdS$_5\times {\rm U(1)}$. A solution of this type in the conformal window
for general $\nc,\nf$ was constructed in Ref.~\cite{Klebanov:2004ya} from the effective action
of six-dimensional non-critical superstrings. Even though this solution is subject to 
order one corrections from higher-derivative terms in the effective action, it does provide the 
required background for the strongly-coupled deformed SQCD. In the string frame the effective 
5D coupling is~\cite{Klebanov:2004ya}
\beq
     \frac{1}{g_5^2} = \frac{3}{4k} N F,
\eeq
where $k$ is the AdS curvature scale, so that in the limit $\nc,\nf\gg 1$, with the ratio $\nc/\nf$ fixed, 
the theory becomes weakly coupled. The AdS$_5$ curvature scale is of order the string scale and 
in the conformal window the radius of the circle $S^1$ satisfies 
$\frac{1}{9} < k R_{S^1} < \frac{2}{9}$.

As a first approximation we can ignore the $S^1$ since its radius remains approximately constant, 
and consider the 5D spacetime $x^M = (x^\mu,z)$ 
with the fifth dimension $z$ compactified on an $S^{1}/Z_2$ interval with 3-branes located at 
$\zuv=k^{-1}\sim \luv^{-1}$ (the UV brane) and $\zir\sim \lir^{-1}$ (the IR brane) where $\zuv/\zir=\lir/\luv$. The AdS$_5$ metric is written using conformal coordinates
\beq
ds^{2}=\frac{1}{(k z)^2}(\eta_{\mu\nu}dx^{\mu}dx^{\nu}+dz^{2}),
\eeq
where $\eta_{\mu\nu} ={\rm diag}(-+++)$ is the 4D Minkowski metric. 

The 5D field content is dictated by the AdS/CFT dictionary. In the 4D electric theory (SQCD)
at strong coupling, corresponding to $F = 3/2 (N+\nu)$,  each operator ${\cal O}(x)$ 
corresponds to a field $\Phi(x,z)$ in the 5D bulk theory. Furthermore, global symmetries of the 
4D theory are interpreted as local gauge symmetries in the bulk. Therefore we assume that the 
5D theory has an $SU(F)_L\times SU(F)_R\times U(1)_B$ gauge symmetry. Note that the $U(1)_R$ 
symmetry is associated with the isometry of the $S^1$. 
We are interested in a deformation corresponding 
to chiral symmetry breaking. Therefore we introduce bulk fields $A_{L\mu}^a$ and $A_{R\mu}^a$,
with $a=1\dots N$ corresponding to the vector currents of the quark fields, and a bifundamental field 
$\Phi^i_j$ with $i,j = 1\dots F$, corresponding to the operator $Q^i\tilde Q_j$ (in the sense that 
the UV VEV of $\p$ determines $m_Q$). The 5D masses $m_5$ 
of the bulk fields are determined by the relation, $m_5^2=(\Delta -p)(\Delta+p-4) k^2$ where $\Delta$ 
is the dimension of the corresponding $p$-form operator~\cite{Gubser:1998bc, Witten:1998qj}. 
For a vector current with $\Delta=3$ this corresponds to massless 5D vector fields. The dimensionality of 
the bulk field corresponding to the squark bilinear $Q\tilde Q$ (whose dimension is $4-{\rm dim}(Q\tilde Q)$)
can be deduced from the $R$-charges of $Q$ in Table~\ref{sqcd0} to be $\Delta=4-3(1-\nc/\nf)$ 
so that $2<\Delta<3 $ inside the conformal window. Likewise the field corresponding to the
quark bilinear has $\Delta=3-3(1-\nc/\nf)$ with $1<\Delta<2 $. The scalar field component of $\Phi$ 
has a 5D mass-squared $m_{5\Phi}^2 = - 3k^2(1-\nc/\nf) (1+3\nc/\nf)$ which approaches 
$ - 3k^2$ at $\nf = \frac{3}{2}\nc$. In the underlying non-critical string theory the vector fields represent 
55 strings propagating on the worldvolume  of the $F$ spacetime filling D5 branes, while $\Phi$ 
represents the open string tachyons on the D5 branes. 

The 5D Lagrangian consists of two parts involving a bulk ${\cal N} =2$ vector supermultiplet 
$(V,\chi)$ containing an ${\cal N}=1$ vector supermultiplet, $V$ and chiral supermultiplet $\chi$, 
and an ${\cal N} =2$ hypermultiplet $(\Phi,\Phi_c)$, containing ${\cal N}=1$ chiral supermultiplets $\Phi,\Phi_c$.
It is given by \cite{Gherghetta:2000qt, Marti:2001iw,Gherghetta:2010cj,Cacciapaglia:2008bi}
\beq
\begin{split}
{\cal S} =\int d^{5}x&\Biggl\{\int d^{4}\theta\, \frac{1}{(kz)^3}\,\left[\p e^{-V}\p^\dagger 
+\p_{c}e^{V}\p_c^{\dagger} + \frac{2}{g_5^2} \left(\partial_5 V - \frac{k z}{\sqrt{2}}(\chi+\chi^\dagger)\right)^2 \right]\\
&+\int d^{2}\theta\, \left[\frac{1}{4g_5^2} W_\alpha W^\alpha+ \frac{1}{(kz)^3}\,\Bigl[\p_{c}
\left(\mathcal{D}_{z}-(\frac{3}{2}-c)\frac{1}{z}\right)\p\right.\\
&\left.+ \delta(z-\zuv)\wuv+\delta(z-\zir)\wir\Bigr]+{\rm h.c.}\right]\Biggr\}\\
= \int d^5x& \sqrt{-g} \left[ - |D \Phi |^2 -m_{5\Phi}^2  |\Phi|^2 -\frac{1}{4g_L^2} F_L^2 
-\frac{1}{4g_R^2} F_R^2+ \dots\right]~,
\end{split}
\label{5daction}
\eeq
where $D_\mu\Phi = \partial_\mu\Phi - i A_{L\mu}\Phi +i A_{R\mu}\Phi$ with $A_{L,R}=A_{L,R}^a t^a$, 
and $c$ is a bulk mass parameter. Note that at the massless level the orbifold breaks the ${\cal N} =2$ supersymmetry down to ${\cal N}=1$ supersymmetry. We shall choose the $\p$ superfield to be even 
under the orbifold action, and $\p_c$ to be odd.

Inserting the $F$-term equations,
\begin{eqnarray}
   F^* & = & \partial_{z}\phi_{c} -\left(\frac{3}{2}+c\right) \frac{1}{z} \phi_{c}-\delta(z-\zuv)\partial_{\p}\wuv
   -\delta(z-\zir) \partial_{\p}\wir \, ,\nonumber \\
    F^*_{{c}} & = & -\partial_{z}\phi+\left(\frac{3}{2}-c\right)\frac{1}{z}\phi
    \label{eq:f-terms}\, ,
\end{eqnarray}
into the equations of motion results in a set of well-known bulk solutions for the components of $\Phi$ 
at arbitrary momentum, $p$. The bulk solutions for the zero modes ($p=0$) correspond simply to setting 
the $F$-terms to zero\cite{Gherghetta:2000qt, Marti:2001iw,Cacciapaglia:2008bi}
\ba
\label{etavev}
\phi (z)&=&\phi(\zuv)\left(\frac{z}{\zuv}\right)^{\frac{3}{2}-c}\, ,\nn\\
\phi_c(z)&=&  \varepsilon(z) {\phi}_c(\zuv^{+})\left(\frac{z}{\zuv}\right)^{\frac{3}{2}+c}\, ,
\ea
where $\varepsilon (z) = $~sign$(z)$ forces $\phi_c$ to be odd and where $\phi_c$ and $\phi$ are the 
scalar components of $\p_c$ and $\p$ respectively. 
Since we are choosing $\p_c$ to be odd under the orbifolding
then we can just set $\phi_c(\zir)=\phi_c(\zuv)=\delta F_c(\zir)=\delta F_c(\zuv)=0$. 
Demanding the vanishing of the delta-function contributions to $F^*$ gives the additional 
constraint 
\beq 
\label{cond1}
\left. \frac{1}{(kz)^{\frac{3}{2}+c}} \frac{\partial W}{\partial \p} \right|_{z=\zuv}=  
- \left. \frac{1}{(kz)^{\frac{3}{2}+c}} \frac{\partial W}{\partial \p} \right|_{z=\zir}\, .
\eeq
Alternatively this condition can be derived by considering boundaries at
$\zuv^+$ and $\zir^-$. Generally the solutions then
have to be consistent with the vanishing of the boundary terms in the variation of the 
action:
\ba
\label{acvary}
\delta S  & = &  \int d^4 x \, 
 \left[ (k z)^{-3} \left(   \delta F_c \frac{\partial W}{\partial \Phi_c} +  \delta F\frac{\partial W}{\partial \Phi} 
\right) \right]_{z=\zuv}
\hspace{-0.5cm}+ \left[ (k z)^{-3} \left(   \delta F_c \frac{\partial W}{\partial \Phi_c} +  \delta F\frac{\partial W}
{\partial \Phi}\right) \right]_{z=\zir}\, 
\nn\\
&& \hspace{1cm}+\frac{1}{2}  \left[ (k z)^{-3} 
 \left(  \delta \phi F_c - \delta \phi_c F - \phi\, \delta F_c + \phi_c\, \delta F \right) \right]^{z=\zuv}_{z=\zir} \, .
\ea
Setting the $\delta F$ component to zero with the solutions of Eq.\eqref{etavev} also gives Eq.\eqref{cond1}.

Using the AdS/CFT relation $\Delta = 2 +\sqrt{4+m_{5\Phi}^2/k^2}$ with $m_{5\Phi}^2 = (c^2+c-15/4)k^2$
we find that $\Delta=5/2+c$. The $\phi$ solution in \eqref{etavev} therefore behaves like 
$\phi(z) \sim z^{4-\Delta}$.
These general solutions have to be matched to whatever VEV $\p$ may acquire due to brane 
interaction terms. In particular we are interested in deformations of the strongly coupled 
SQCD by the addition of a quark mass term and a consequent explicit breaking of the chiral symmetry. 
In the 5D description this corresponds to specifying the UV boundary condition of the bifundamental 
field $\Phi^i_j$, such that 
\beq
\label{boundary}
      \frac{1}{\sqrt{k}(kz)^{4-\Delta}}\Phi^i_j\Big|_{\zuv} = (m_Q)^i_j~.
\eeq

\subsection{The gravitational dual descriptions of Seiberg duality}
 
 We now connect this AdS picture with the underlying Seiberg duality, in 
particular identifying the possible configurations of the bulk and brane theories, the location
 of the various degrees of freedom involved, and the brane superpotentials $\wuv$ and $\wir$.

 It is natural to suppose in the RS1 context that the composite degrees of freedom (i.e. the weakly coupled 
 magnetic theory) should be placed on the IR brane. Note that the magnetic $\SU{n}$ gauge fields are forced to appear there because they must be 
purely {\em emergent} degrees of freedom -- they can have no bulk presence since that would imply that a global symmetry of the strongly coupled theory has become a gauge theory at low energies, contradicting the theorem of 
Ref.\cite{Weinberg:1980kq}. 
We have seen that in order for the underlying strongly coupled theory to have the required mass term, $W\supset m_Q Q\tilde{Q}$, there must be an additional  bulk meson $\p$ whose UV boundary value, by the bulk/boundary correspondence, acts as a source field for the operator ${\cal{O}}=Q\tilde{Q}$ through 
the interaction \begin{equation} \label{source} W \supset  \p Q\tilde{Q}.\eeq  
In accord with $\welec$ of the electric Seiberg dual, the value of $\p $ on the 
UV brane should be determined by $m_Q$ as specified in Eq.\eqref{boundary}. 
Note that as described in Ref.\cite{Cacciapaglia:2008bi}, the holographic correspondence in the supersymmetric  theory 
appears as an interaction in the superpotential on the UV boundary.
Hence the correspondence is between a source and current superfield so that,
in particular, the scalar operator $\cal{O}$ of the CFT couples to the $F$-term of the source superfield.
\begin{figure}[p]
\begin{center}
  \begin{picture}(420,148) (131,-190)
    \SetWidth{2.7}
    \SetColor{Blue}
    \Line(186,-44)(186,-188)
    \SetColor{Red}
    \Line(438,-60)(438,-172)
    \Text(296,-130)[lb]{\Huge{\Black{$\mbox{$\Phi$, $V_\nf$}$}}}
    \Text(445,-106)[lb]{\Large{\Black{$\mbox{$\varphi$, $q$, $\tilde{q}$, $v_\nnc$}$}}}
    \Text(171,-106)[lb]{\Large{\Black{$\mbox{$\eta$}$}}}
    \Text(110,-130)[lb]{{\Black{$\mbox{$\wuv=$}$}}}
    \Text(445,-130)[lb]{{\Black{$\mbox{$\wir=\, q\pp\tilde{q} -  m_\varphi \p \pp$}$}}}
    \Text(113,-148)[lb]{{\Black{$\mbox{$m_\eta \, \eta (\p-\p_0)$}$}}}
    \SetWidth{1.0}
    \SetColor{Black}
    \Bezier(192,-44)(304,-92)(400,-92)(432,-92)
    \Bezier(192,-188)(304,-140)(400,-140)(432,-140)
  \end{picture}
\end{center}
\caption{\label{setup} \em The first configuration for Seiberg duality in the large $\nc$ limit. This corresponds to the ``usual'' case in which  there is a massless meson.The bulk gauge symmetry is $\SU{\nf}_L\times\SU{\nf}_R\times \Uo_B$. The unbroken symmetry on the IR-brane, $\SU{\nf_2}_{L}\times\SU{\nf_2}_{R}\times \Uo_{B}$, corresponds to the light quarks in the electric dual. The upper ${\nf_1}\times{\nf_1}$ flavour block of $\Phi_0$ has VEVs of $\sqrt{k} m_0$, and is consequently Higgsed out of the low energy theory by the $F_\pp=0$ condition, at a scale 
$\langle \hat{q}\hat{\tilde{q}} \rangle = (\sqrt{k}m_\pp) \lir $.}
\end{figure}
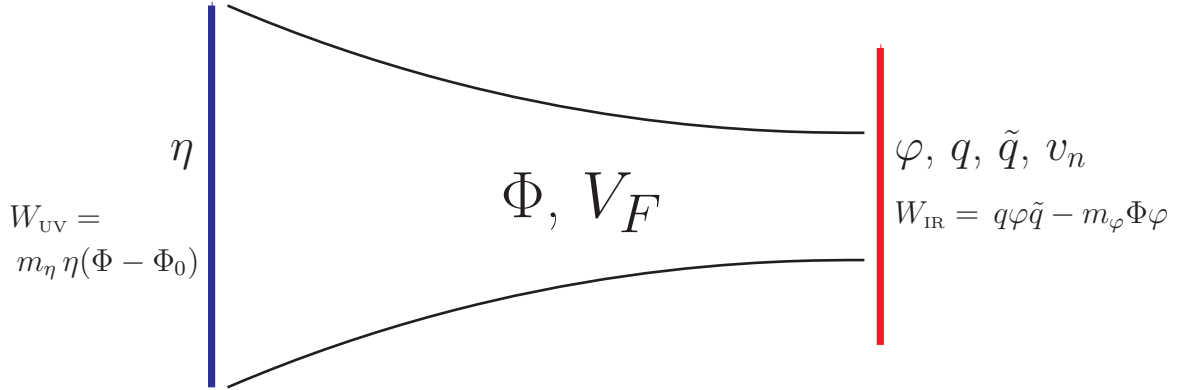

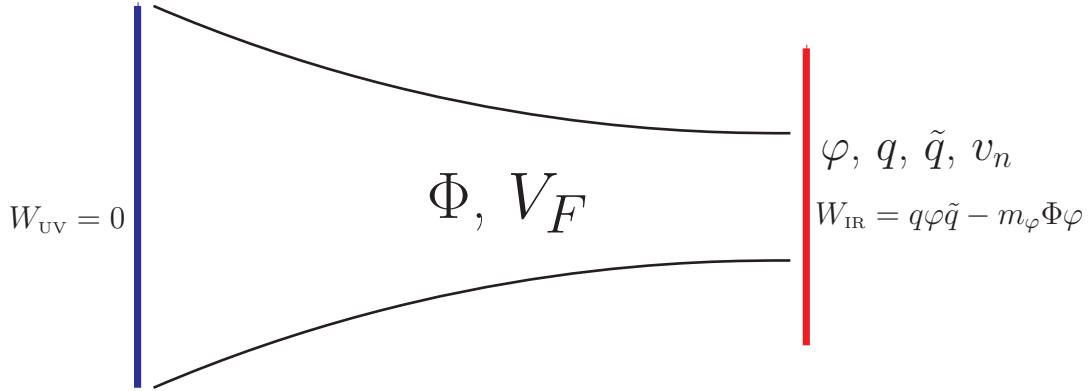
\begin{figure}[p]
\begin{center}
  \begin{picture}(420,148) (131,-190)
    \SetWidth{2.7}
    \SetColor{Blue}
    \Line(186,-44)(186,-188)
    \SetColor{Red}
    \Line(438,-60)(438,-172)
    \Text(296,-130)[lb]{\Huge{\Black{$\mbox{$\Phi$, $V_\nf$}$}}}
    \Text(445,-106)[lb]{\Large{\Black{$\mbox{$\pp$, $q$, $\tilde{q}$, $v_\nnc$}$}}}
    \Text(138,-130)[lb]{{\Black{$\mbox{$\wuv=0$}$}}}
    \Text(443,-130)[lb]{{\Black{$\mbox{$\wir= q\pp\tilde{q}- m_\pp \p \pp$}$}}}
    \SetWidth{1.0}
    \SetColor{Black}
    \Bezier(192,-44)(304,-92)(400,-92)(432,-92)
    \Bezier(192,-188)(304,-140)(400,-140)(432,-140)
  \end{picture}
\end{center}
\caption{\label{setup2} \em The second configuration for Seiberg duality in the large $\nc$ limit, when there is no massless meson, but a field $\p$ and a coupling $\p Q\tilde{Q}$ in the electric superpotential. 
As in Figure~\ref{setup}, the bulk meson is the source field for the composite 
operator ${\cal{O}}=Q\tilde{Q}$ (i.e. its UV value corresponds to the quark masses). 
}
\end{figure}

The fact that $m_Q$ is associated with the dynamical degree of freedom $\p$ in the AdS picture implies that we should base the rest of the structure on the {\em dual-of-the-dual} electric theory. To see this, let us briefly return to 
4D Seiberg duality and rename the meson in the magnetic theory $\eta\equiv  Q\tilde{Q}/\Lambda$, so that the superpotential 
is $\wmag = \eta q\tilde{q}$. (We shall henceforth drop the $h$-couplings unless we are dealing with them specifically, and shall assume them to be of 
order unity. We shall also for the moment 
set the dynamical scales of the electric and magnetic theories to be degenerate, $\Lambda$.) 
On performing the dual of this theory one arrives at an alternative (dual-of-the-dual) electric theory with meson 
$\p \equiv  -q\tilde{q}/\Lambda$ and superpotential coupling 
\beq
\label{welecp}
\welec' = \p Q\tilde{Q} -\Lambda \p \eta \, .
\eeq  
The first term of Eq.\eqref{welecp} is of course precisely the required source term of Eq.\eqref{source}. 
(The minus sign comes from the matching of dynamical scales in the electric and magnetic theories, as 
described below in Section 3.3.) One of the checks of Seiberg duality is that this theory flows to the same IR 
physics as the theory we first started with. Indeed both $\p$ and $\eta$ have masses of order $\Lambda $. 
Upon integrating them out one finds $\eta =  Q\tilde{Q} /\Lambda$, leading to the same spectrum and zero superpotential as the original theory. However note that we may also choose to keep all the degrees of 
freedom and dualise (yet again) to a second {\em magnetic} theory. Now the mesons 
$\p$ and $\eta$ are to be treated as elementary and a new composite meson $\pp\equiv Q\tilde{Q}/\Lambda$ is introduced: the superpotential is  
\begin{equation}
\label{wmagp}
\wmag'=\, q\pp\tilde{q} +  \Lambda \p \pp - \Lambda \p \eta   \, .
\eeq
 Integrating out $\p $ and $\pp_-=\frac{1}{\sqrt{2}}(\eta-\pp)$ leaves us with 
 the magnetic spectrum and superpotential ($\wmag =  q\pp_+\tilde{q} $ where 
 $\pp_+=\frac{1}{\sqrt{2}}(\eta+\pp)$) as required. It also identifies the two mesons $\pp$ and $\eta$. 
 
 \newpage 
 
With this in mind, let us now return to the gravitational dual. 
 The 4D discussion above implies two possible 
 gravitational dual configurations in the large $\nc$ limit: 
 
\begin{itemize}

\item  {\bf {\em Configuration 1 -- a massless meson $\pp$}}: In this case, 
because we have a dynamical $\p$, we must add a 
new elementary meson $\eta$ on the UV brane that has the same quantum numbers as $\pp$. 
The superpotential on the UV brane must be 
\beq
\wuv =  m_\eta (\p-\p_0) \eta \, ,
\eeq   
where $\p_0 = m_Q \sqrt{k}(k\zuv)^{4-\Delta} $ fixes the UV VEV of $\p$ through the $F_\eta =0$ equation of motion. 
The superpotential on the IR brane is of the form 
\begin{equation}
\label{wir}
\wir=\, q\pp\tilde{q} - m_\varphi \p \pp \, .
\end{equation}
Next let us discuss scales. First note that, using $\Delta_\pp= \frac{3}{2}-c$, the purely 4D relation in Eq.\eqref{eq:lamir}  gives  
\beq
\label{zoob}
\lir = m_0 \left( \frac{\zuv}{\zir} \right)^{\frac{1}{2}+c}\, ,
\eeq
so that the factor that naturally warps the physical scales in this set-up is $\left( {\zuv}/{\zir} \right)^{\frac{1}{2}+c}$.
Now, by comparison with Eq.\eqref{wmagp}, $m_\varphi\sim m_\eta$ are identified with the compositeness scale, but not directly. Indeed because $\p$ is a 5D field $m_\pp$ and $m_\eta$ have mass dimension $1/2$.
Furthermore canonical normalisation of the fields on the IR brane leads to the physical Yukawa coupling being 
of order unity, hence the normalised brane fields are $\hat{q}=q/(k\zir)$,  $\hat{\tilde{q}}=\tilde{q}/ (k\zir)$ and 
$\hat{\pp}=\pp/(k\zir)$.
Thus once the fields are canonically normalised, the term $m_\varphi $ gives a physical mass that is also
naturally warped down by a factor $({\zuv/\zir})^{\frac{1}{2}+c}$ compared to that generated by $m_\eta$.
(For the most part we will allow $m_\pp$ and $m_\eta$ to be free parameters.) 
The relation of $m_\pp$ and $m_\eta$ to the compositeness scale can be determined by the Higgsing in the
$\SU{\nf_1}_L\times \SU{\nf_1}_R$  flavour-block of the magnetic theory
corresponding to the heavy quarks of the electric dual: upon canonically normalising the magnetic quarks and using 
Eqs.\eqref{etavev}, \eqref{boundary} and \eqref{zoob}, the $F_\pp=0$ condition 
for $\wir$ gives  
\beq
\langle \hat{q}\hat{\tilde{q}} \rangle = (k\zuv)^{4-\Delta} (\sqrt{k}m_\pp) \lir \, .
\eeq
Thus, since $\lir $ is the physical mass of the heavy quarks in the electric dual, we can identify 
\[
(k\zuv)^{4-\Delta}\sqrt{k}m_\pp 
\] as 
the compositeness scale, and set its value to be $\sim \lir$. Furthermore the boundary condition 
in Eq.\eqref{cond1} correctly equates the two canonically normalised mesons $\hat{\pp}$ and $\hat{\eta}$ 
and their masses as
\beq
\label{foobar}
\left( \frac{\zuv}{\zir} \right)^{\frac{1}{2}+c} m_\pp \hat{\pp} = m_\eta \hat{\eta} \, .
\eeq
This identifies the heavy mode as being mostly $\eta$ and the orthogonal massless mode as being mostly $\pp$,
provided that $c>-\frac{1}{2}$.  Note that if $m_\pp=0$, then $\eta=0$
 and we recover the standard Seiberg picture in which the massless meson is identified 
 entirely as $\varphi$, and $\eta$ is integrated out. 
The set-up for this configuration is as shown in Figure \ref{setup}. 

\item {\bf\em Configuration 2 -- no massless meson}: In the 4D theory this 
corresponds to having just the source term in the electric superpotential without the $\eta$ field: 
\beq
\label{welecp2}
\welec =  \p Q\tilde{Q} .
\eeq  
Upon dualizing one finds a magnetic superpotential 
\begin{equation}
\label{wmagp2}
\wmag=\, q\pp\tilde{q} +  \Lambda \p \pp   \, .
\eeq
Thus both $\p$ and $\pp$ gain a mass of order $\Lambda$ and may be integrated out of the 
low energy theory, leaving no mesons and no superpotential in the magnetic theory. The gravitational dual 
has a bulk field for every CFT operator so \mbox{$\p\leftrightarrow  Q\tilde{Q}$} must
 be there by the bulk/boundary correspondence. Hence the bulk configuration, shown in Figure \ref{setup2}, 
 must be as before but with no additional $\eta$ meson.
 For the boundary superpotentials in the gravitational dual we have 
 \beq
\wuv =  0 \, ,
\eeq  
and 
\begin{equation}
\label{wir2}
\wir=\, q\pp\tilde{q} - m_\varphi \p \pp \, .\end{equation}
 
\end{itemize}
As the elementary $\eta$ and composite $\varphi$ mix, an obvious possible extension
is to generalize the picture and have  just a single bulk field $\varphi$.
However by the bulk/boundary correspondence, the VEV of this field on
the UV brane would be identified as the  source for an operator in the underlying CFT (i.e. the electric theory). 
This operator would have to have the same flavour charges as $\Phi \sim q\tilde{q}$
but such an operator is not readily available in the electric Seiberg dual (a possible exception 
being the case $\nf=\nc+1$, when the magnetic quarks can be written as electric baryons).
Hence although the physics would be similar to that of configuration 1,
the direct link to strongly coupled 4D Seiberg duals would be lost.

\subsection{Two caveats}
 
 \label{gauging}
 
 In the context of AdS/CFT the flavour symmetries of our strongly coupled 4D theory are weakly gauged, and 
their RG behaviour (in particular the possibility of Landau poles) is then important. Let us first consider 
this from a 4D point of view, by returning to the electric theory with $\nf = \frac{3}{2}(\nc +\nu )$ with 
$\nc \gg \nu$. The $\sunc$ gauge groups are inside the conformal window, but  the (now gauged) flavour groups, $\sunf_L\times\sunf_R$, see $\hat{n}_f=\nc/2$ flavours and $\hat{n}_c=\nf$ colours. However the 
flavour groups are anomalous and must be cancelled. According to the anomaly matching 
procedure of \thooft (see below), the flavour anomalies of the electric and magnetic descriptions should both be 
cancelled by the same spectator sector. In the present case the coefficients of the $\sunf_L^3$ and $\sunf_R^3$ 
anomalies are $-\nc$ and $\nc$ respectively.
Thus the spectator sector could be an additional $\nc$ fundamentals of $\sunf_L$ and $\nc$ antifundamentals of $\sunf_R$.
The total number of $\sunf$ flavours is then $\hat{n}_f=\nc $ and colours is $\hat{n}_c=\nf=\frac{3}{2}(\hat{n}_f +\nu )$.
Since $\hat{n}_f < \hat{n}_c$ one would conclude that a vacuum does not exist for these theories.
However one may add additional states with masses around $\lir$ which are charged only under the flavour symmetry. As far as the Seiberg duality is concerned these states are gauge singlets, and clearly their additional contribution to $\hat{n}_f$ 
can bring the $\sunf_L\times\sunf_R$ flavour symmetry within its conformal window as well. 

What if one instead has a theory 
where the flavour symmetries are in the IR-free phase, with effectively $\hat{n}_f > 3\hat{n}_c$? 
(We shall encounter an example later.) In that case the flavour symmetries would hit a 
Landau pole in the UV and another Seiberg duality would be required. From a 4D viewpoint the original $\sunc$ would in turn become IR-free and lead to a duality cascade involving the flavour symmetry as well (and most likely a duality wall would ensue). This can be averted by instead adding massive {\em gauge states} around the 
scale $\lir$. These come in ${\cal N}=2$ vector multiplets; they can be decomposed as an ${\cal N}=1$ vector and an ${\cal N}=1$ chiral superfield both in the adjoint. Hence a single massive gauge multiplet contributes $2\hat{n}_c$ to the beta function coefficient and such states can again bring the flavour groups inside the conformal window. 

In short therefore, one expects that even from an entirely 4 dimensional point of view, 
additional massive states can tame the behaviour of the flavour symmetries above the scale $\lir$
should we choose to gauge them. In the 5D AdS picture such states correspond to Kaluza-Klein modes
of the bulk gauge fields. The net effect of the coupling to the CFT is that 
the RG behaviour of the flavour groups becomes logarithmic~\cite{Pomarol:2000hp,ArkaniHamed:2000ds}. 
A bound results on the contribution that the flavoured degrees of freedom in the CFT make to the beta function
in order to avoid Landau poles in the flavour couplings below the UV scale~\cite{Reece:2010xj}:
\beq
\label{boundy}
\bcft \lesssim \frac{2\pi}{\alpha(\lir)}\frac{1}{\log{(\luv/\lir)}}\, .
\eeq
When taking the Veneziano limit we necessarily have $\bcft\sim \nc$ and hence this translates into an 
an upper bound on $\nc$ which depends on $\alpha (\lir)$. For RS1 type scenarios where the flavour symmetries are 
identified with SM gauge symmetries, and where $\lir \sim $1~TeV this bound is very strict, $\nc \lesssim {\cal O}(10)$. 
In other cases, for example when the bulk gauge symmetries are not SM gauge symmetries (with the latter being 
either emergent symmetries located on the IR brane or having very small values in the IR), or in the ``little RS'' scenario~\cite{Davoudiasl:2008hx}, the bound can be greatly relaxed. 

These bounds are generally equivalent to a {\em lower} bound on the curvature of any would-be gravitational dual~\cite{Reece:2010xj}. Indeed 
the radius of AdS is typically given by $R_{\mbox{\tiny AdS}}\sim \ell_s N^\delta $ where for example 
$\delta = 1/4$ for $AdS_5\times S_5$, so that the bound on $N$ immediately translates into a bound on 
$R_{\mbox{\tiny AdS}}$. In the sub-critical construction of Ref.\cite{Klebanov:2004ya} which we shall 
be invoking, we have $\delta=0$ and 
then the curvature is quite large anyway, $R_{\mbox{\tiny AdS}}\sim \sqrt{6} \ell_s $. This is in line with the 
general expectation of Ref.~\cite{Bigazzi:2005md}.

This then is the first caveat about the validity of this framework that one must bear in mind. The second 
is related, and concerns the interpretation of the dynamical scales of the field theory. 
In particular, the rightmost panel in 
Figure \ref{fig:flows} is  the 4D field theoretical situation whose large $\nc $ limit we are supposed to be 
approximating. But, in the figure, both electric and magnetic theories have their Landau poles 
around the scale $\lir$, and look to be strongly coupled there: why then are we entitled to place 
a weakly coupled theory on the IR brane? The answer can be found in the matching relation for dynamical scales 
in Seiberg duality which still permits the Landau poles of the electric and magnetic descriptions to be in different places. 

To see this, we can return to the 4D theories and ``blow up'' the running around the scale $\lir$ in order 
to examine what 
happens there when we match the two descriptions. Let us treat the dynamical scales 
of the electric theory ($\Lambda$) and magnetic theory ($\bar\Lambda$) more carefully, and 
temporarily reinstate the coupling $h$ which we have been setting to one above.
We will need the matching relation for Seiberg duality, which is 
\beq
\label{foo}
{\bar{\Lambda}}^{\bar{b}} \Lambda^b = (-)^{\nf-\nc} \hat{\Lambda}^{b+\bar{b}}\, , 
\eeq 
expressed in a basis where the quarks of both theories are canonically normalised, and where the magnetic 
superpotential is written 
\beq
\wmag = \frac{1}{\hat{\Lambda}} Q\tilde{Q} q\tilde{q} + m_Q Q\tilde{Q}\, .
\eeq
In Eq.\eqref{foo}, $b=3\nc-\nf$ and $\bar{b}=3\nnc -\nf$ are the one-loop beta-function coefficients so that 
\ba
\frac{4\pi}{\alpha(t)} &=& b (t-t_\Lambda)\, ,\nn \\
\frac{4\pi}{{\bar\alpha}(t)} &=& {\bar b} (t-t_{\bar \Lambda})\, ,
\ea
where $t=\log E$. 
For an SQCD theory in the free magnetic phase we have $b>0$ and ${\bar{b}}<0$, so that we get a perturbative overlap of the electric and magnetic 
descriptions when $\hat{\Lambda}> \Lambda > {\bar \Lambda}$. In the interval between $\Lambda$ and $\bar\Lambda$ both descriptions are in principle valid.  The relation is illustrated graphically in Figure~\ref{fig:matching}. 
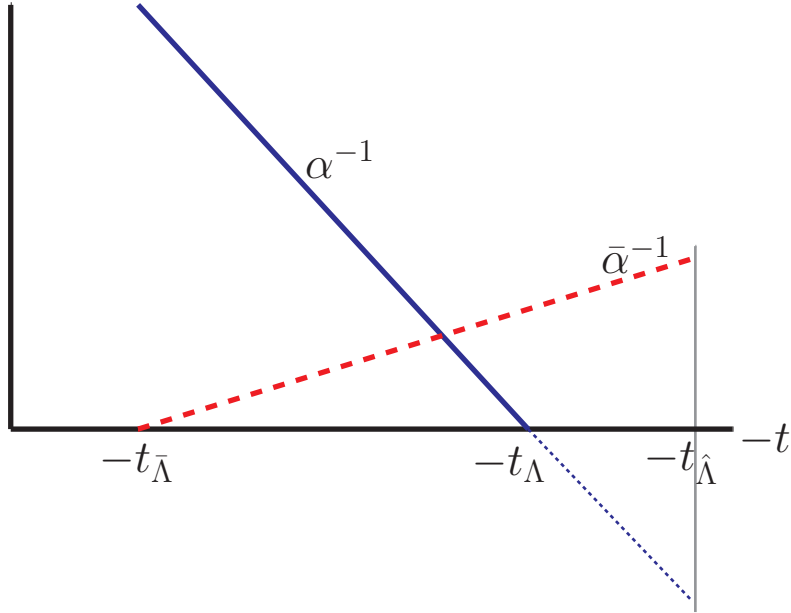
\begin{figure}
\begin{center}
  \begin{picture}(368,225) (115,-95)
    \SetWidth{1.0}
    \SetColor{Gray}
    \Line(418,38)(418,-100)
    \SetWidth{2.0}
    \SetColor{Black}
    \Line(160,129)(160,-31)
    \Line(432,-31)(160,-31)
    \SetWidth{2.0}
    \SetColor{Blue}
    \Line(355,-31)(208,129)
    \Text(335,-51)[lb]{\Large{\Black{${ -t_\Lambda}$}}}
    \Text(436,-41)[lb]{\Large{\Black{$-t$}}}
    \Text(194,-51)[lb]{\Large{\Black{$-t_{\bar \Lambda}$}}}
    \Text(400,-52)[lb]{\Large{\Black{$-t_{\hat{\Lambda}}$}}}
    \Text(384,28)[lb]{\Large{\Black{${\bar \alpha}^{-1}$}}}
    \Text(272,64)[lb]{\Large{\Black{$\alpha^{-1}$}}}
    \SetColor{Red}
    \Line[dash,dashsize=4.4](208,-31)(416,33)
    \SetWidth{1.0}
    \SetColor{Blue}
    \Line[dash,dashsize=1.5](355,-31)(416,-95)
  \end{picture}
\end{center}
\caption{\label{fig:matching}\em 
Representation of the scale matching of Eq.\eqref{foo}. The parameter $t=\log(E)$, so that the IR is to the right in 
the diagram. The scale $\hat{\Lambda}$ is formally where $\alpha=-{\bar \alpha}$, as indicated by the extension of the $\alpha^{-1}$ line below the axis. }
\end{figure}
The scale $\hat{\Lambda}$ is unknown, nevertheless as shown later in Section~\ref{secISS} 
(see Eq.\eqref{wmag2}) one can identify  
\ba
h \pp &=& \frac{Q\tilde{Q}}{\hat{\Lambda}} \nn \\
h \mu_\pp^2 &=& m_Q \hat{\Lambda}\, ,
\ea
where $\pp $ is the canonically normalised meson. On dimensional grounds one expects the K\"ahler potential 
to have terms of the form 
\beq
K \supset \frac{ {Q\tilde{Q}} {Q^\dagger\tilde{Q}^\dagger} }{\kappa^2 \Lambda^2 }\equiv \pp \pp^\dagger\, ,
\eeq
where $\kappa$ is some unknown coefficient. The coupling, $h$, is related to this coefficient as 
\beq
\label{gam}
h=\frac{\Lambda \kappa}{\hat{\Lambda}}\, .
\eeq
Thus in addition to $m_Q$ there are three free (or rather unknown) parameters which we can choose to be $h$, $\Lambda$ and 
$\bar{\Lambda }$ (with $\hat\Lambda$ and $\kappa $ being determined by Eqs.\eqref{foo} and \eqref{gam}).

It is clear from this discussion what must happen near the IR brane when we have the more complicated situation described 
above, namely a conformal interval terminated by quark mass terms that cause the theory to enter the free magnetic phase 
below their mass, $\lir$. In that case the behaviour shown in Figure \ref{fig:confmatching} is perfectly acceptable. 
Above $\lir$ the coupling does not run. Below the scale $\lir$ the electric coupling quickly becomes non perturbative. But the magnetic theory can be made arbitrarily weakly coupled in this region since $\bar\Lambda $ is a free parameter, irrespective 
of how strongly coupled is the electric theory. 
\begin{figure}
\begin{center}
  \begin{picture}(368,188) (115,-95)
    \SetWidth{0.1}
    \SetColor{Gray}
    \GBox(335,-68)(352,71){0.956}
     \SetWidth{2.0}
    \SetColor{Black}
    \Line(160,92)(160,-68)
    \Line(432,-68)(160,-68)
    \SetWidth{2}
    \SetColor{Blue}
    \Line(352,-68)(336,-52)
    \Text(343,-86)[lb]{\Large{\Black{${ -t_\Lambda}$}}}
    \Text(436,-82)[lb]{\Large{\Black{$-t$}}}
    \Text(194,-90)[lb]{\Large{\Black{$-t_{\bar \Lambda}$}}}
    \Text(305,-90)[lb]{\Large{\Black{$-t_{\lir}$}}}
    \Text(384,-9)[lb]{\Large{\Black{${\bar \alpha}^{-1}$}}}
    \Text(222,-50)[lb]{\Large{\Black{$\alpha^{-1}$}}}
    \SetColor{Red}
    \Line[dash,dashsize=4.4](337,-29)(416,-4)
    \SetColor{Blue}
    \Line(336,-52)(208,-52)
    \SetWidth{0.1}
    \SetColor{Violet}
    \Line[dash,dashsize=2](336,-52)(208,92)
    \SetWidth{2.0}
    \SetColor{Red}
    \SetWidth{0.1}
    \Line[dash,dashsize=2.5](335,-31)(206,-68)
  \end{picture}
\end{center}
\caption{\label{fig:confmatching}\em 
As in Figure~\ref{fig:matching} but with the theory entering a conformal phase above the scale $\lir$. Above this scale the 
couplings do not run. Below the scale $\lir$ the electric coupling quickly becomes non perturbative. The ``IR brane'' is represented by the 
grey strip in the interval between $\lir$ and $\Lambda$. The magnetic theory is weakly coupled in this region.}
\end{figure}
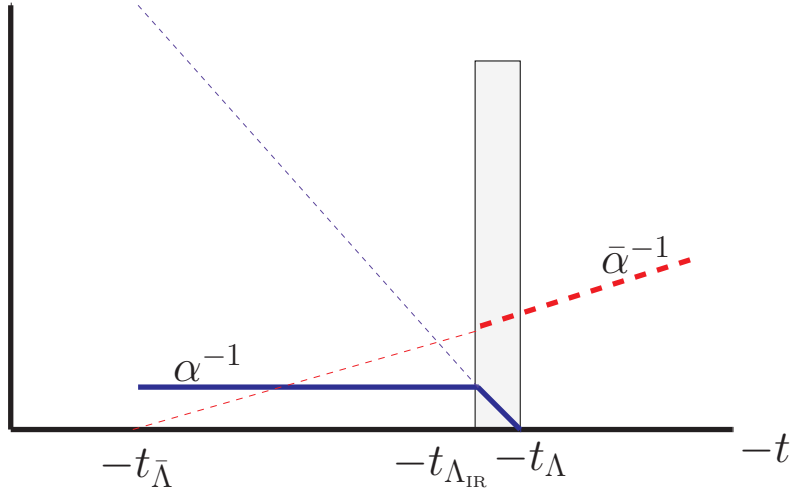

\subsection{Flow in the Klebanov-Maldacena solution}

Klebanov and Maldacena considered the 6d non-critical string action and 
found an AdS$_5\times$~$S_1$ solution in the presence of $\nf$ space filling D5 and 
anti-D5 branes and with $\tilde{\nc}={\cal O} (\nc)$ units of RR flux.
Despite the fact that these solutions are relatively strongly curved, it is argued 
that they are a qualitative approximation to the gravitational dual of SQCD. However 
for our 5D AdS approximation to be valid it is important that the radius of $S_1$ is 
small and constant. This is indeed the case, and actually the more general solutions of this system 
can be seen to exhibit some of the features of the flows we see in the gauge theory. 
The ansatz for the 6d metric in the string frame is 
\beq 
ds^2 =  
e^{2f }dx^2 + dr^2 + e^{2g}d\theta^2 \, .
\eeq
Thus the radius of curvature of the $S_1$ is $R_{S_1}=\sqrt{\alpha'} e^g$. 
The evolutions of the scalars is given by 
\ba
g' &=& \frac{{\tilde{\nc}}}{\nf} e^{-g}- \frac{\nf}{{\tilde{\nc}}} e^{g}-\frac{{\tilde{\nc}}}{2} e^{\phi-g}\nn \\
\phi' &=& - \frac{\nf}{{\tilde{\nc}}} e^{g}+{{\tilde{\nc}}} e^{\phi-g}\nn \\
f' &=& \frac{{\tilde{\nc}}}{2} e^{\phi-g}\, ,
\ea
and where $\phi$ is the dilaton.
For constant $\nf$, the solutions exhibit two typical kinds of behaviour. The choice $e^{2g}=\frac{2}{3}\frac{{\tilde{\nc}}^2}{\nf^2}$ and 
$\nf e^\phi =\frac{2}{3}$ is the AdS$_5\times$~$S_1$ solution with constant $R_{S_1}$. If we perturb away from this solution 
it is straightforward to see that the theory will flow to a linear dilaton behaviour in the UV (i.e. as $r\rightarrow  \infty $). The 
dilaton behaves as $\phi \rightarrow\phi_0-\frac{\nf}{\tilde{\nc}}r$ with the theory becoming weakly coupled,  
while for generic starting values of $g$ we have $e^g\rightarrow \frac{\tilde{\nc}}{\nf} $ (asymptoting as a tanh function), and 
$f\rightarrow const$. This behaviour is clearly seen in Figure~\ref{fig:flows2}a where we show $g(r)$ and $\phi(r)$. The flat region 
is an AdS$_5\times$~$S_1$ ``slice'', while the left of the plot (the UV) is the linear dilaton region. 

It is interesting to look at the solutions when $\nf $ is itself a function of $r$. This is the gravitational equivalent 
of integrating out heavy quarks below their mass on the gauge side: in this case it would correspond to a recombination of the 
D5/anti-D5 flavour branes in the bulk. We can mimic this by adjusting $\nf $ by hand at some value of $r$. The resulting behaviour 
is shown in Figure ~\ref{fig:flows2}b. A change in flavour perturbs the AdS$_5\times$~$S_1$ solution to one with a different 
$R_{S_1}$ curvature and dilaton, with the AdS$_5$ curvature remaining the same. There is an obvious analogy with Figure~\ref{fig:flows}. 
\vspace{1cm}
\begin{figure}[h]
\begin{centering}
\includegraphics[angle=0,scale=.6]{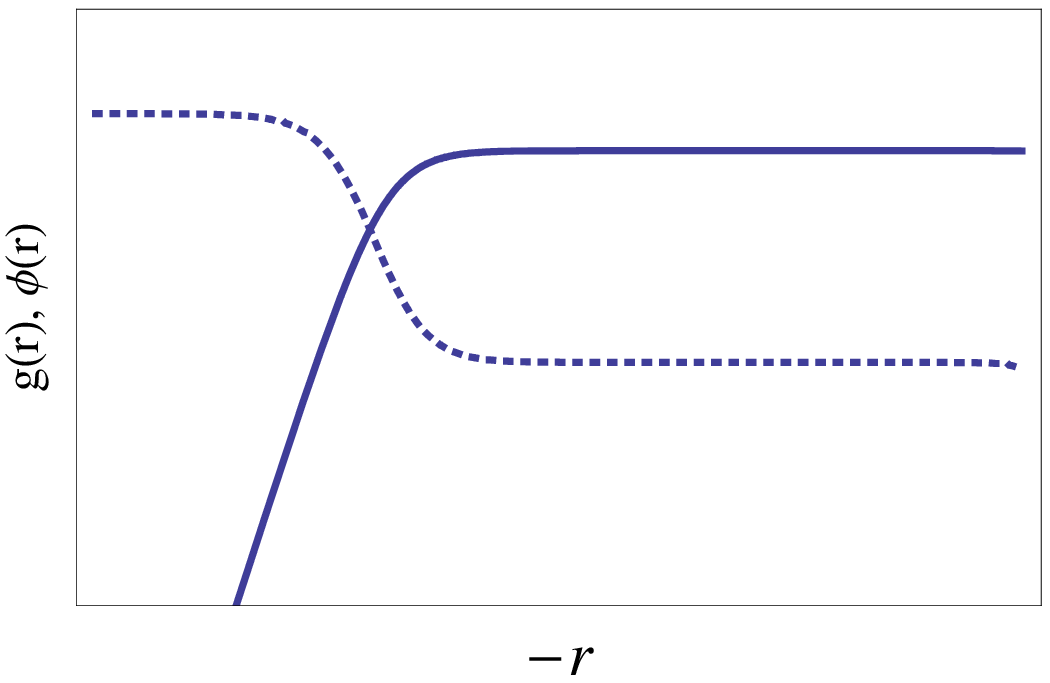}
\includegraphics[angle=0,scale=.6]{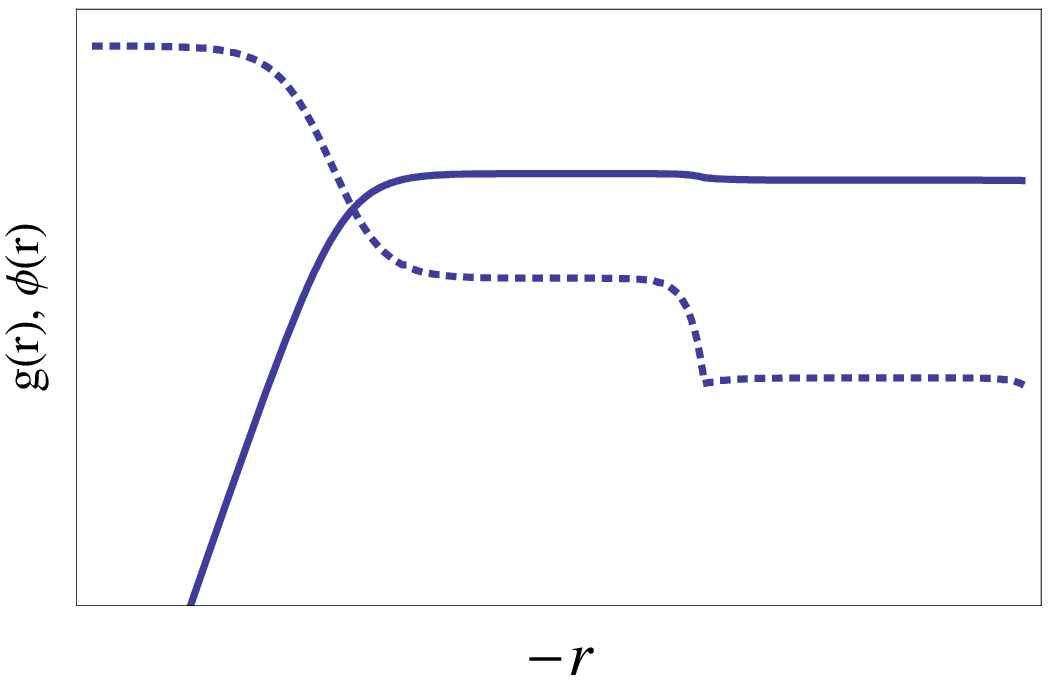}
\par\end{centering}
\caption{\it Flow in the Klebanov-Maldacena solution. Left: the warping $g(r)$ (dotted) and dilaton $\phi(r)$ (solid)
for fixed $\nf$. Right: $g(r)$ (dotted) and dilaton $\phi(r)$ (solid) for an $r$-dependent $\nf$.} 
\label{fig:flows2}
\end{figure}

\subsection{Anomaly matching in RS1}

As we have already mentioned the electric and magnetic theories of Tables~\ref{sqcd0} and \ref{sqcd0-1} are anomalous 
in the flavour symmetries: for example both the the $\sunf^3_L$ and  $\sunf^3_R$ anomalies are non-zero.  In a
purely 4D setting the `t~Hooft anomaly 
matching idea asserts that such flavour anomalies should be the same in both theories; the reasoning is that if one were to gauge the flavour 
symmetries one would have to add a spectator sector charged only under the flavour symmetries in order to cancel these 
anomalies. 
However these new sectors would be blind to the $\sunc$ gauge group and so insensitive to 
its behaviour. (By analogy, the lepton sector of the Standard Model does not change when QCD confines.) 
That both electric and magnetic descriptions have the same flavour anomalies is one of the most powerful tests 
of Seiberg duality in 4D. 

Now however the situation has changed: the flavour symmetries {\em really are} gauged in the bulk. 
What, therefore, becomes of anomaly matching -- is it even consistent with our picture of AdS as the 
large $\nc $ limit of Seiberg duality? (See also \cite{Abel:2010uw}.) In order to address this question 
let us consider what the spectator sector has to be. Because anomalies do not exist in 5D, the 4D anomalies 
are associated with contributions localised at the branes \cite{ArkaniHamed:2001is,Scrucca:2001eb,Barbieri:2002ic,GrootNibbelink:2002qp}. They can be written
\begin{equation}
\sqrt{-g} {\cal A}^a(x,z)={\cal A}^a_{\mbox{\tiny UV}}(x)\,\delta(z-\zuv)+{\cal A}^a_{\mbox{\tiny IR}}(x)\,\delta(z-\zir)\, ,
\end{equation}
where for $i=UV, IR$
\begin{equation}\label{4danomaly}
{\cal A}^a_i=\frac{n_i}{24\pi^2}\varepsilon^{\mu\nu\kappa\lambda}\tr\left[T^a\,\partial_\mu \left(A_\nu\partial_\kappa A_\lambda+\frac{1}{2} A_\nu A_\kappa A_\lambda\right)\right].
\end{equation}
Here the gauge fields can be associated with the 4D zero modes up to an overall normalisation
since their bulk profiles are flat.
The constants $n_i$ receive the following contributions. A left-chiral fundamental fermion localised on the
UV (IR) brane contributes $\nuv=1,\nir=0$ ($\nuv=0,\nir=1$).
The contribution from a massless left-chiral bulk fermion is $\nuv=\nir=1/2$.
On an $S^1/(Z_2\times Z_2')$ orbifold heavy 
modes with parities $(+-)$ and $(-+)$ can give localised 
anomalies that are equal and opposite on 
the branes. The 
contribution from a fundamental fermion with parity $(+-)$ is $\nuv=-\nir=1/2$ and
from a fermion with parity $(-+)$, it is $\nuv=-\nir=-1/2$. Finally there is 
a contribution from a 5D bulk Chern--Simons term
which also gives equal and opposite
gauge anomaly terms on the branes.
In a consistent gauge invariant theory the summed contributions  should vanish on both boundaries. 

Now let us interpret the bulk and brane fields in terms of Seiberg duality and discuss the 
roles that they would play in anomaly matching. Fields 
on the IR brane clearly belong
 to the magnetic theory -- we shall call this contribution $\nir^{\mbox{\tiny (mag)}}$.
Fields in the bulk have both a composite and an elementary component -- they are therefore 
present in both electric and magnetic theories, and moreover they are singlets of both the magnetic 
gauge group (because that lives on the IR-brane only) and the electric gauge group. They can therefore 
be thought of as part of the spectator sector. They contribute the same anomaly at each brane:
\[ 
\nuv^{\mbox{\tiny (bulk)}}=\nir^{\mbox{\tiny (bulk)}}= \frac{1}{2} n^{\mbox{\tiny (bulk)}}\, ,
\]
where $n^{\mbox{\tiny (bulk)}}$ is their total contribution to the 4D anomalies. 
Finally the Chern-Simons and possibly heavy mode contributions are 
\[ 
- \nuv^{\mbox{\tiny (CS)}}= \nir^{\mbox{\tiny (CS)}}= n^{\mbox{\tiny (CS)}}\, .
\]

As we have said the total contribution must vanish at each boundary. Thus at the IR boundary we have 
to satisfy 
\beq
n^{\mbox{\tiny (mag)}} + \frac{1}{2}n^{\mbox{\tiny (bulk)}} +  n^{\mbox{\tiny (CS)}} \, = \, 0\, .
\eeq
 The interpretation in terms of \thooft anomaly matching of the 4D field theories is that
 a spectator sector added to cancel the anomalies of the magnetic theory would have total anomaly 
\beq
n^{\mbox{\tiny (spec-mag)}} =  \frac{1}{2} n^{\mbox{\tiny (bulk)}} + n^{\mbox{\tiny (CS)}} \, .
\eeq

On the other hand the UV brane has total anomaly 
\beq
\nuv^{\mbox{\tiny (brane)}} + \frac{1}{2} n^{\mbox{\tiny (bulk)}} - n^{\mbox{\tiny (CS)}} \, = \, 0\, ,
\eeq
where $\nuv^{\mbox{\tiny (brane)}}$ are the contributions from whatever degrees of freedom 
are localised on the UV brane. In order to cancel anomaly contributions here we require
\beq
 \nuv^{\mbox{\tiny (brane)}}+ n^{\mbox{\tiny (bulk)}} =  n^{\mbox{\tiny (spec-mag)}} \, . 
\eeq
But the left hand side of this equation is the total contribution to the flavour anomaly 
coming from the elementary degrees of freedom that couple to the CFT. This is nothing other than
the spectator sector as seen by the underlying electric theory, $n^{\mbox{\tiny (spec-elec)}}$. Thus the anomaly matching condition (i.e. $n^{\mbox{\tiny (spec-elec)}}=n^{\mbox{\tiny (spec-mag)}}$) is satisfied: the same spectator sector can serve to cancel anomalies in both the electric and magnetic theories. 
Note that the essential features of the bulk physics required for this to work are 
that a) the contributions from the bulk modes are the same at each 
brane and b) the contributions from the Chern-Simons terms and the heavy modes are equal and opposite. 

As an explicit example consider the anomalies when there is a single bulk meson field $\Phi$. In total this contributes $-\nf$ 
to the $\sunf^3_L$ anomalies (and the negative to the $\sunf_R^3$ anomaly), with one half appearing at each boundary. 
The magnetic theory on the IR brane 
has a contribution of $\nc$. Hence a
would-be spectator sector (including $\Phi$)  that could cancel the anomaly would have to contribute $-\nc $.
However
the Chern-Simons contribution has to be $ \frac{1}{2}\nf- \nc$ at the IR-brane. On the UV brane 
the net anomaly from $\Phi$ and the Chern-Simons term is $\nc- \nf = -\nnc$ so we have to add additional 
elementary fields here that contribute $\nnc$ to the $\sunf^3_L$ anomaly. From a 4D perspective 
the total contribution from the elementary degrees of freedom (i.e. the spectator sector) is then this plus the contribution from $\Phi$, giving the same $-\nc$ that we required for the magnetic theory.
In this example the elementary sector could be an additional $\nnc$ fundamentals of $\sunf_L$ and  
$\nnc$ antifundamentals of $\sunf_R$.

\section{The MSSM in RS1}

We now turn to applications of the proposed ``large $\nc$'' limit of Seiberg duality, the first 
being an explicit realisation of the Randall-Sundrum (RS1) idea, that the particles of the 
Standard Model arise as composite degrees of freedom on the IR brane of a slice of AdS. 
Of course in their original proposal Randall and Sundrum were interested in protecting 
the Higgs mass in non-supersymmetric scenarios. Since we are working in Seiberg duality 
we will consider the ${\cal N}=1$ supersymmetric MSSM instead. (This model is morally similar 
to the super-technicolour models of Ref.~\cite{Antola:2010nt}.)

In Table~\ref{mssm} we show the MSSM theory. This is the theory that we wish to put on the 
IR brane of a slice of AdS. The advantage of our approach is that we can establish the set-up 
working mostly in 4D field theory. Following the discussion of the previous sections we will 
do this by deriving the MSSM as the magnetic dual of an electric theory. Additional states will then be 
integrated into the latter in order to make the electric phase conformal above some mass scale, and then the 
``large $\nc$'' limit will be taken. This last step requires some particle content in the magnetic theory that can be adjusted -- this content is an arbitrary number of $n_h$ 
Higgs pairs, with $n_h=1$ corresponding to the usual MSSM. The gauge group that we will dualize is
$\SU{2}_L\simeq \Sp{1}$. 

We assume for concreteness that the Higgs fields, being in vector-like pairs, have a generic range of 
mass terms, and that the lightest Higgs is the only one that ends up with a VEV (i.e. this is the one that plays the 
role of the usual MSSM Higgs).  One can envisage more complicated cases in which the electroweak breaking 
is distributed amongst the Higgs pairs. At low scales, the theory runs precisely as the MSSM, with 
an asymptotically free $\SU{3}$ and a mildly positive beta function for $\Sp{1}$: the 
$\Sp{1}$ group sees an effective flavour number $f_{\Sp{1}}=7$, and the 
beta function coefficient is given by  
 $\bar{b}_{\Sp{\mmc}}=3(\mmc+1)-f_{\Sp{\mmc}}$, so that $\bar{b}_{\Sp{1}}=-1$ as usual. 
 At energy scales above the masses of the additional Higgs fields, those states can be ``integrated in'' to the 
 theory and start to contribute to the running as well. Eventually we have 
$f_{\Sp{1}}=6+n_h$. The $\SU{3}$ running continues unchanged but the beta 
 function of $\Sp{1}$ becomes more negative, eventually reaching $\bar{b}_{\Sp{1}}=-(1+n_h)$. Therefore 
 we can expect the theory to reach a Landau pole at some scale $\Lambda$, above which an electric description takes over.
 
 In order to get to an electric description we use the Seiberg duality for $\Sp{\mmc}$ groups; the 
 electric gauge group is $\Sp{\mc}$ where 
 \beq
  \mc = f_{\Sp{\mmc}}-(\mmc+2) = 3+n_h\, .
 \eeq
 The content of the theory is shown in Table~\ref{sqcd0-3}. 
In this section we will use the convention that additional Higgs-like states with $R$-parity $R_p=1$ are denoted either with $\p$ or $\pp$, with a suffix to indicate which SM field they resemble charge-wise. 
 
The $R$-parity charges in the electric theory allow the following superpotential (we set all couplings to one and suppress generation indices to avoid clutter):
 \beq
 \label{welec1}
 \welec= 
\p_E L L + \p_D Q L+\p_{D^c}\, Q Q + \p_S\, Q Q  +\p_\sigma H_U H_D + \mu_{\sigma}^2 
\p_\sigma\, .
 \eeq 
 This theory has a number of features that we should comment on. 
 First the charges of the Higgses, left-handed quarks and left-handed leptons are the opposite of 
 the charges of their magnetic counterparts. This is because  the 
 gauge groups other than $\Sp{1}$ are playing the role of ``flavour'' in the Seiberg duality, and as usual 
 those charges must be reversed. Second, note that 
 the right handed fields of the magnetic theory have no counterpart in the electric theory: in fact those fields are 
 composite bound state ``mesons'' of the electric theory. In addition there are some new states charged 
 under $\SU{3}$ and $\Uo_Y$. These can be identified as the composite ``mesons'' of the magnetic theory (the term ``meson'' of course always referring to mesons of the Sp groups):
\ba
\p_E &\leftrightarrow & ll \nn \\
\p_D &\leftrightarrow & lq \nn \\
\p_{D^c} &\leftrightarrow & q_{[i} q_{j]} \nn \\
\p_S &\leftrightarrow & q_{\{i} q_{j\} } \nn\\
\p_\sigma &\leftrightarrow & h_{u} h_{d} \, .
\ea
The trilinear terms in Eq.\eqref{welec1} are simply the superpotential terms associated with that identification (c.f. Eq.\eqref{wmag0}).
Finally the linear singlet term in Eq.\eqref{welec1}  corresponds to the Higgs mass terms of the magnetic theory: the 
masses are $\sim \mu_\sigma^2/ \Lambda $ where $\Lambda $ is the Landau pole scale of the magnetic theory, hence we require that 
\beq
\mu_\sigma < \Lambda \, .
 \eeq
\begin{table}[htp]
\centering{}\begin{tabular}{|c|c|c|c|c|}
\hline 
\multicolumn{1}{|c|}{$ $} & $\SU{3}$ & $\Sp{1}$  & $\U1_{Y}$ & $R_p$\tabularnewline
\hline 
\hline 
$q_i$ & $\fund$ & $\fund$ & $\frac{1}{6}$ &  $-1$\tabularnewline
\hline
$l_i$ & $1$ & $\fund$ & $-\frac{1}{2}$ &  $-1$\tabularnewline
\hline
$n_h\times h_u$ & $1$ & $\fund$ & $\frac{1}{2}$ &  $1$\tabularnewline
\hline
$n_h\times h_d$ & $1$ & $\fund$ & $-\frac{1}{2}$ &  $1$\tabularnewline
\hline
\hline
$e^c_i$ & $1$ & $1$ & $1$ &  $-1$\tabularnewline
\hline
$\nu^c_i$ & $1$ & $1$ & $0$ &  $-1$\tabularnewline
\hline
$d^c_i$ & $\afund$ & $1$ & $\frac{1}{3}$ &  $-1$\tabularnewline
\hline
$u^c_i$ & $\afund$ & $1$ & $-\frac{2}{3}$ &  $-1$\tabularnewline
\hline
\end{tabular}\caption{\emph{The MSSM spectrum augmented by $n_h-1$ additional massive Higgs pairs. 
The index $i=1\ldots 3$ is the usual generation index.\vspace{0.4cm}}
\label{mssm}}
\end{table}

\begin{table}[h!]
\centering{}\begin{tabular}{|c|c|c|c|c|}
\hline 
\multicolumn{1}{|c|}{$ $} & $\SU{3}$ & $\Sp{3+n_h}$  & $\U1_{Y}$ & $R_p$\tabularnewline
\hline 
\hline 
$Q_i$ & $\afund$ & $\fund$ & $-\frac{1}{6}$ &  $-1$\tabularnewline
\hline
$L_i$ & $1$ & $\fund$ & $\frac{1}{2}$ &  $-1$\tabularnewline
\hline
$n_h\times H_U$ & $1$ & $\fund$ & $-\frac{1}{2}$ &  $1$\tabularnewline
\hline
$n_h\times H_D$ & $1$ & $\fund$ & $\frac{1}{2}$ &  $1$\tabularnewline
\hline
\hline
$3\times \p_E$ & $1$ & $1$ & $-1$ &  $1$\tabularnewline
\hline
$6\times \p_{D^c}$ & $\afund$ & $1$ & $\frac{1}{3}$ &  $1$\tabularnewline
\hline
$3\times \p_S$ & $\symm$ & $1$ & $\frac{1}{3}$ &  $1$\tabularnewline
\hline
$9\times \p_D$ & $\fund$ & $1$ & $-\frac{1}{3}$ &  $1$\tabularnewline
\hline
singlets $\p_\sigma$ & $1$ & $1$ & $0$ &  $1$\tabularnewline
\hline
\end{tabular}\caption{\emph{The electric dual theory.}
\label{sqcd0-3}}
\end{table}
\begin{table}[h!]
\centering{}\begin{tabular}{|c|c|c|c|c|}
\hline 
\multicolumn{1}{|c|}{$ $} & $\SU{3}$ & $\Sp{1}$  & $\U1_{Y}$ & $R_p$\tabularnewline
\hline 
\hline 
$q_i$ & $\fund$ & $\fund$ & $\frac{1}{6}$ &  $-1$\tabularnewline
\hline
$l_i$ & $1$ & $\fund$ & $-\frac{1}{2}$ &  $-1$\tabularnewline
\hline
$n_h\times h_u$ & $1$ & $\fund$ & $\frac{1}{2}$ &  $1$\tabularnewline
\hline
$n_h\times h_d$ & $1$ & $\fund$ & $-\frac{1}{2}$ &  $1$\tabularnewline
\hline
\hline
$e^c_i$ & $1$ & $1$ & $1$ &  $-1$\tabularnewline
\hline
$\nu^c_i$ & $1$ & $1$ & $0$ &  $-1$\tabularnewline
\hline
$d^c_i$ & $\afund$ & $1$ & $\frac{1}{3}$ &  $-1$\tabularnewline
\hline
$u^c_i$ & $\afund$ & $1$ & $-\frac{2}{3}$ &  $-1$\tabularnewline
\hline
\hline
$3\times \pp_{{e}^c}$ & $1$ & $1$ & $1$ &  $1$\tabularnewline
\hline
$6\times \pp_d $ & $\fund$ & $1$ & $-\frac{1}{3}$ &  $1$\tabularnewline
\hline
$3\times \pp_{s^c}$ & $\bar{\symm}$ & $1$ & $-\frac{1}{3}$ &  $1$\tabularnewline
\hline
$9\times \pp_{d^c}$ & $\afund$ & $1$ & $\frac{1}{3}$ &  $1$\tabularnewline
\hline
singlets $\pp_\sigma$ & $1$ & $1$ & $0$ &  $1$\tabularnewline
\hline
\hline
$3\times \p_E$ & $1$ & $1$ & $-1$ &  $1$\tabularnewline
\hline
$6\times \p_{D^c}$ & $\afund$ & $1$ & $\frac{1}{3}$ &  $1$\tabularnewline
\hline
$3\times \p_S$ & $\symm$ & $1$ & $\frac{1}{3}$ &  $1$\tabularnewline
\hline
$9\times \p_D$ & $\fund$ & $1$ & $-\frac{1}{3}$ &  $1$\tabularnewline
\hline
singlets $\p_\sigma$ & $1$ & $1$ & $0$ &  $1$\tabularnewline
\hline
\end{tabular}\caption{\emph{Flowing down from the electric theory. 
This magnetic theory is arrived at by dualizing the electric theory of Table~\ref{sqcd0-3}. One finds mass terms in the 
superpotential for the states in the 
last two blocks, and they can be integrated out to arrive at the MSSM spectrum of Table~\ref{mssm}.}
\label{sqcd0-bar}}
\end{table}
It is instructive to confirm that the electric theory we have just described does indeed flow to the 
MSSM (augmented by extra Higgs fields) of Table~\ref{mssm}. 
(As we have presented it, this is in fact nothing other than the test that the dual-of-a-dual gives back the original theory \cite{Intriligator:1995au}.) 
Upon performing the Seiberg duality of the electric theory 
we find the spectrum of Table~\ref{sqcd0-bar}. As we have already said, the block of right-handed MSSM fields are 
bound state ``mesons'' of the electric theory. The identification is 
\ba
e_i^c     &\leftrightarrow & L_i H_D      \nn \\
\nu_i^c &\leftrightarrow & L_iH_U       \nn \\
d^c_i    &\leftrightarrow & Q_{i} H_D    \nn \\
u_i^c    &\leftrightarrow & Q_{i} H_U    \, .
\ea
The third block of states are also composite ``mesons'' of the magnetic theory:
\ba
3\times \pp_{{e}^c} &\leftrightarrow & LL \nn \\
6\times \pp_d &\leftrightarrow & Q_{[ i}Q_{ j]} \nn \\
3\times \pp_{s^c} &\leftrightarrow & Q_{\{ i} Q_{j\} } \nn \\
9\times \pp_{{d}^c} &\leftrightarrow & QL \nn \\
\mbox{singlets } \pp_\sigma &\leftrightarrow & H_{U} H_{D} \, .
\ea
The superpotential of {\em this} theory is 
\ba
 \label{wmag2}
 \wmag &=& 
 \Lambda (
 \p_E \pp_{{e}^c} +  \p_D \pp_{{d}^c}+\p_{D^c} \pp_d + \p_S \pp_{s^c}  +\p_\sigma \pp_\sigma ) + \mu_{\scriptstyle \sigma}^2 \p_\sigma\nn \\
 && + e^c lh_d + \nu^c lh_u + d^c q h_d + u^c q h_u + \pp_\sigma h_u h_d \, .
 \ea 
The first set of terms are masses of order $\Lambda$ so these fields can be integrated out, whereupon we 
recover the original spectrum of Table~\ref{sqcd0-bar} and the MSSM superpotential 
\ba
 \label{wmag3}
 \wmag &=& 
 \frac{ \mu_{\scriptstyle \sigma}^2}{\Lambda} h_u h_d  + e^c lh_d + \nu^c lh_u + d^c q h_d + u^c q h_u \, .
 \ea 
The first term gives as required the set of masses for the extra Higgs fields (including one for the 
lightest field that would correspond to the usual ``$\mu$-term'' of the MSSM). The trilinear terms are 
the usual set of Yukawa couplings: in Seiberg duality we have the added bonus that these interactions necessarily arise 
because the right-handed fields are composite mesons. 
Naturally there is then the question of how one could explain the hierarchy that we observe in these
interactions. This is outside the scope of the present work but is surely an interesting topic for future study. 

Having found a candidate electric dual for the MSSM, all that remains is to take the 
large $\mc$ limit and propose a gravitational dual. 
In this case we first add into the electric theory the required states to bring it into the 
conformal window of the Sp group. The free magnetic window is given by 
\ba
  \mc+2 < f_{\Sp{\mc}}\leq \frac{3}{2} (\mc+1)\, ,
 \ea
and the conformal window by
\ba
  \frac{3}{2}(\mc+1) < f_{\Sp{\mc}}< {3} (\mc+1)\, .
 \ea
Initially (i.e. for the theory of Table~\ref{sqcd0-3}) the electric theory has  $\mc=n_h+3$ and 
$f_{\Sp{\mc}}=n_h+6$ which, as expected, places it within the free magnetic range for any value of $n_h> 0$.
 
Now let us add some additional Higgs states but with masses $\lir$. Below $\lir$ these fields are 
integrated out and the theory enters the free magnetic phase described above; thus as discussed in 
section~\ref{UVcompletion} it is the mass $\lir$ which generates the ``IR-brane''. 
We will add an extra $n'_h$ pairs of Higgses $H'_U$ and $H'_D$ so that above $\lir$ 
we have $f_{\Sp{\mc}}=n'_h+n_h+6$.  In order to bring the electric theory into the conformal window above 
$\lir$ we can define a parameter $\nu$ such that  
\beq
n'_h = \frac{1}{2}(n_h+3\nu)\, ,
\eeq 
so that 
\beq
  f_{\Sp{\mc}}=  \frac{3}{2}(\mc+1+\nu)\, .
\eeq
We are then free to take the large $\mc$ limit. In this case sending $n_h \rightarrow \infty$ but keeping $\nu={\cal O}(1)$
formally gives a parametrically strongly coupled conformal $\Sp{M}$. Of course none of this changes the magnetic theory which remains the MSSM with some extra Higgses.

Finally we should discuss the running of the SU(3) and $\Uo_Y$ groups. The electric theory has a large gauge group 
$\Sp{n_h+3}$. From the point of view of the $\SU{3}$ group this provides a large number of additional flavours. 
In fact the effective number of flavours contributing to the $\SU{3}$ beta function is 
\beq 
f_{\SU{3}}=3(n_h+11)\, .
\eeq
Clearly the $\SU{3} $ group is in principle now highly IR-free in the large $n_h$ limit (and even 
when $n_h=1$). 
Usually one would expect the theory to exhibit some sort of cascade behaviour above $\lir$ with $\SU{3}$ hitting a Landau pole. 
In fact the situation is the one outlined in Section~\ref{gauging}. In a conformal phase these beta functions are tamed
by additional massive modes charged only under $\SU{3}$ (or $\Uo_Y$) appearing at the scale $\Lambda$. 
From the RS point of view these states appear automatically as the low-lying Kaluza-Klein modes of the bulk 
gauge and matter fields. As explained, the end result is a logarithmic running and a bound on the value of $\mc$. 

Having made this caveat we can now propose the entire 
gravitational dual of this theory at large M.   The
set-up is shown in Figure \ref{setup-mssm}. The $\SU{2}_L$ group is emergent and so must appear on the IR brane. From the perspective of the strongly coupled theory, the remaining gauge symmetries are 
flavour symmetries and have to appear in the bulk. From Section \ref{5ddual}, 
it is clear that all the magnetic ``quarks'' denoted generically by $\bf q$ (i.e. ${\bf q}=q,l,h_u,h_d$) and every low energy ``matter'' meson of the Sp groups with R-parity $-1$, denoted by ${\boldsymbol {\pp}_-}$ (i.e. ${\boldsymbol \pp_-}= e^c, \nu^c,d^c, u^c$), also appear on the IR brane. 
Moreover as in the field theory $\wir \supset {\bf q \boldsymbol {\pp}_-{q}}$ on the IR-brane gives the 
required MSSM superpotential terms. 

However there is a further modification required in the gravitational dual because there are bulk fields corresponding to every composite operator of the CFT. Therefore 
the bulk contains not only the original $R_p=+1$ fields as above denoted generically as ${\boldsymbol \p}_+$,
\beq
{\boldsymbol \p}_+=
\left\{
\begin{array}{lll}
\p_E &\leftrightarrow & ll \nn \\
\p_D &\leftrightarrow & lq \nn \\
\p_{D^c} &\leftrightarrow & q_{[i} q_{j]} \nn \\
\p_S &\leftrightarrow & q_{\{i} q_{j\} } \nn \\
\p_\sigma &\leftrightarrow & h_{u} h_{d} \, ,
\end{array}
\right.
\eeq
but also the $R_p=-1$ bulk fields that did not appear in the field theory, denoted generically by ${\boldsymbol \p}_-$:
\beq
{\boldsymbol \p}_-=
\left\{
\begin{array}{lll}
E &\leftrightarrow & lh_d \nn \\
\nu &\leftrightarrow & lh_u \nn \\
{D} &\leftrightarrow & q h_d \nn \\
U &\leftrightarrow & q h_u  \, .
\end{array}
\right.
\eeq
As we saw in section~\ref{5ddual} the two types of meson are distinguished by the 
fact that the latter are in configuration 1, with a corresponding set of matter fields (i.e. ${\boldsymbol{\eta}}_-= \hat{e}^c, 
\hat{\nu}^c,\hat{d}^c, \hat{u}^c$) coupling to ${\boldsymbol \p}_-$ on the 
UV brane, whereas the former with no corresponding  ${\boldsymbol{\eta}}_+$
fields on the UV brane, are in configuration 2.
The superpotentials are then given by 
\ba
\wuv & = & m_\eta {\boldsymbol{\eta}}_- {\boldsymbol \p}_- \, , \nn \\
\wir  & = & {\bf q}  {\boldsymbol \pp}_-  {\bf q} + {\bf q}{\boldsymbol \pp}_+{\bf q} - m_\varphi {\boldsymbol \p}_-  {\boldsymbol \pp}_- -  m_\varphi {\boldsymbol \p}_+{\boldsymbol {\pp}_+}\, ,
\ea
where the trilinear couplings automatically contain all the terms consistent with $R$-parity conservation.
Recall that as in the field theory all the mesons  without a UV counterpart (i.e. the ${\boldsymbol \pp}_+$'s) 
are massive and can be integrated out of the low energy theory, whereas those {\em with} a UV counterpart 
(i.e. the ${\boldsymbol \pp}_-$'s) leave a light linear combination of ${\boldsymbol \eta}_-$ and 
${\boldsymbol \pp}_-$ in the low energy theory. Since the wave-function of the ${\boldsymbol \p}_-$ can be warped, the remaining light states could be mostly ${\boldsymbol \pp}_-$ or mostly ${\boldsymbol \eta}_-$. 
Hence the final low energy right-handed fields, ${e^c}', {{\nu}^c}',{{d}^c}', {{u}^c}'$ can naturally have different 
degrees of compositeness, while everything charged under $\SU{2}_L$ must necessarily be entirely composite.
\begin{figure}[h]
\begin{center}
  \begin{picture}(420,148) (131,-190)
    \SetWidth{2.7}
    \SetColor{Blue}
    \Line(186,-44)(186,-188)
    \SetColor{Red}
    \Line(438,-60)(438,-172)
    \Text(225,-124)[lb]{\Large \Black ${\boldsymbol\p}_+$ ${\boldsymbol\p}_-$ 
    $V_{\mbox{       \small SU(3)$_c\times$ U(1)$_Y$}}$}
    \Text(445,-96)[lb]{\Black\small ${\bf q} $ $\equiv$ left-matter+higgs}
    \Text(445,-116)[lb]{\Black\small ${\boldsymbol \pp}_-  $ $\equiv$ comp. right-matter  } 
    \Text(445,-134)[lb]{\Black\small ${\boldsymbol \pp}_+  $ $\equiv$ charged heavy states } 
    \Text(445,-76)[lb]{\Black\large $ v_{\mbox{\tiny SU(2)$_L$}}$ } 
    \Text(115,-106)[lb]{\small{\Black{$\mbox{${\boldsymbol \eta}_-$ $\equiv$ elem.   }$}}}
    \Text(125,-120)[lb]{\small{\Black right-matter   }}
    \Text(103,-150)[lb]{\small {\Black{$\mbox{$\wuv=m_\eta {\boldsymbol \eta}_- {\boldsymbol\p}_-$}$}}}
    \Text(443,-160)[lb]{\small \Black{$\wir= {\bf q}  {\boldsymbol \pp}_-  {\bf q} + {\bf q}{\boldsymbol \pp}_+{\bf q}  $}}
    \Text(467,-177)[lb]{\small \Black $- m_\pp {\boldsymbol \p}_-  {\boldsymbol \pp}_- -  m_\pp {\boldsymbol \p}_+ {\boldsymbol \pp}_+ $}
     \SetWidth{1.0}
    \SetColor{Black}
    \Bezier(192,-44)(304,-92)(400,-92)(432,-92)
    \Bezier(192,-188)(304,-140)(400,-140)(432,-140)
  \end{picture}
\end{center}
\caption{\label{setup-mssm} \em The configuration for the MSSM in RS1 that naturally arises by considering Seiberg duality of $\SU{2}_L\simeq \Sp{1}$. The left-handed matter and Higgs fields are identified as ``quarks'' of the $\Sp{1}$, while a linear combination of the composite ``mesons'' and elementary fields form the light right-handed matter fields.\vspace{0.5cm}
}
\end{figure}
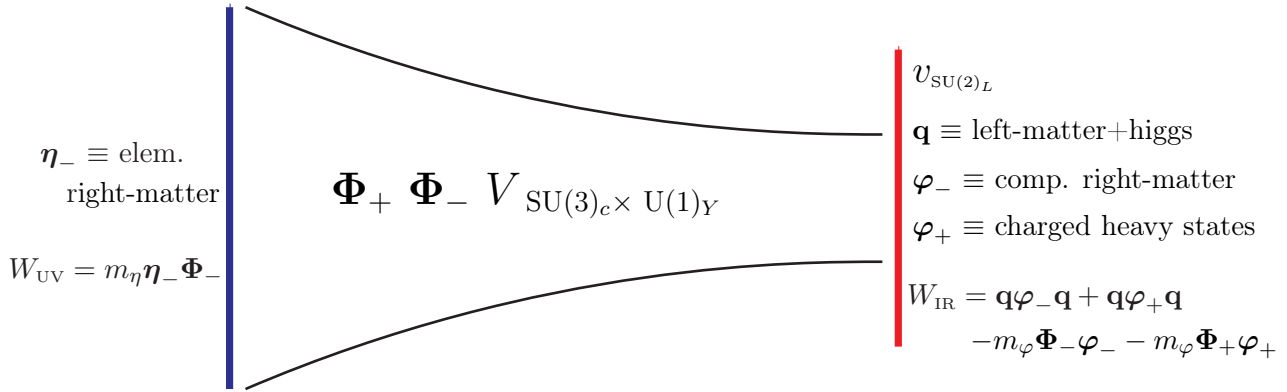

\section{General gauge mediation, simplified}

\vspace{-0.3cm}

\subsection{ ISS}
\label{secISS}

\vspace{-0.2cm}

Our second application is to supersymmetry breaking on the IR brane and its mediation. 
The ``large $\nc $'' limit of Seiberg duality will clearly yield a version of the metastable 
supersymmetry breaking mechanism of Intriligator, Seiberg and Shih \cite{ISS}, but in an RS1 
configuration similar to those discussed in Refs. \cite{Dudas:2007hq,Abel:2010uw}. (A string 
configuration that corresponds to this case was presented in Ref.\cite{Murthy:2007qm}.
This case is morally similar to the metastable superconformal models of Ref.\cite{Amariti:2010sz}.) 

The supersymmetry breaking is a feature of the magnetic theory and so one expects it to appear on the IR-brane. Thus proposals for gauge-mediation that were discussed in the context of RS1 should also be applicable to our strongly coupled configuration. One particular application that we would like to revisit is gauge mediation with gaugino masses that are dominant over scalar ones. In the context of extraordinary gauge mediation~\cite{Cheung:2007es} this corresponds to increasing the ``effective number of messengers'', and a region of 
parameter space that na\"ively corresponds to strong coupling. Calculable and 
explicit models have long been known in the context of extra dimensional  
models \cite{Mirabelli:1997aj,Kaplan:1999ac,Chacko:1999mi,Gherghetta:2000kr,Csaki:2001em,Chacko:2004mi,Gherghetta:2010cj, McGarrie:2010kh}.
Interest has been revived recently in 4D models that can achieve the same kind of screening of scalar
mass contributions in for example Refs.~\cite{McGarrie:2010kh,Green:2010ww,McGarrie:2010qr,Sudano:2010vt}.
Here we shall be using the large $\nc$ limit of the simple perturbative gauge mediation model of Ref.\cite{MN} 
in order to achieve the same effect.

First let us look at the ISS supersymmetry breaking sector and briefly review the model for 
comparison. Ref.\cite{ISS} worked in the free-magnetic phase $\frac{3}{2}\nc \geq \nf \geq \nc+1 $ 
and noted that the classical superpotential $\wmag$ in Eq.\eqref{wmag}  is of the O'Raifeartaigh
type. Supersymmetry breaking occurs because of the so-called rank condition:
\begin{equation}
F_{\pp_{j}^{i}}= \tilde{q}^{j}q_{i}-\mu_\varphi^{2}\,\delta_{i}^j=0\, ,
\end{equation}
can only be satisfied for a rank-$\nnc$ submatrix of the $F_{\pp}$ where $i,j$ are flavour indices.
The height of the potential at the metastable minimum is then given
by \begin{equation}
V_{+}(0)=\nc\,|\mu_\varphi^{4}|\, ,\end{equation}
where for ease of notation we are again setting the coupling $h=1$. 
The supersymmetric minima in the magnetic theory are located by allowing
$\pp$ to develop a vev. The $q$ and $\tilde{q}$ fields
acquire masses of $\langle \pp\rangle$ and can be integrated out,
whereupon one recovers a pure $\sun$ Yang-Mills theory with a nonperturbative
contribution to the superpotential of the form 
\begin{equation}
\wir^{\mbox{\tiny (dyn)}}=\nnc\left(\frac{\mbox{det}\pp}{\bar{\Lambda}^{\nf-3\nnc}}\right)^{\frac{1}{\nnc}}.
\end{equation}
This leads to $\nc$ nonperturbatively generated SUSY preserving minima at 
\begin{equation}
\label{phivev}
\langle \pp_{i}^{j}\rangle=\mu_\varphi\,\epsilon^{-3+2\frac{\nf}{\nc}}\,\delta_{i}^{j}\, ,
\end{equation}
where $\epsilon=\mu_\varphi/\bar{\Lambda}$, in accord with the Witten
index theorem. The minima can be made to appear far from the origin
if $\epsilon$ is small and $3\nc>2\nf$, the condition for the magnetic
theory to be IR-free. The positions of the minima are bounded by the
Landau pole such that they are always in the region of validity of the macroscopic
theory.

Now for the holographic version. Following the discussion in Section~\ref{UVcompletion}, we work down
from the electric theory. In contrast to Ref.\cite{ISS} we begin in the conformal window but with the global
flavour symmetry explicitly broken by relevant mass-terms as
{\small \beq
\sunf_{L}\times\sunf_{R}\times\Uo_{B}\times\Uo_{R} \rightarrow  
\SU{\nf_1}_{D}\times\SU{\nf_2}_{D}\times \SU{\nf_3}_{D}\times  \Uo_{B}  \, ,
\eeq }
where $\nf_1+\nf_2+\nf_3=\nf$, and where we choose 
\ba
&\frac{3}{2}\nc < \nf \leq {3}\nc &\, \nn\\
& \nc+1 \leq (\nf_2+\nf_3) \leq \frac{3}{2}\nc \, . &
\ea
We also add to the electric spectrum a superfield 
$\Phi$  that transforms as an adjoint under the $\SU{\nf_2}_{D}$ symmetry. 
The superpotential of the electric theory is (we set all Yukawa couplings to one)
\beq
 \label{welec2}
 \welec= m_Q Q \tilde{Q}+ Q \Phi \tilde{Q} - \mu_{\scriptstyle \Phi}^2 \Phi\, ,
 \eeq 
where 
\beq
m_Q= \left( 
\begin{array}{lll} 
m_1 {\bf 1}_{\nf_1\times\nf_1} & {\bf 0} &  {\bf 0} \nn\\
 {\bf 0}  & m_2 {\bf 1}_{\nf_2\times \nf_2} & {\bf 0} \nn \\
 {\bf 0}  &   {\bf 0} & m_3 {\bf 1}_{\nf_3\times \nf_3}  \nn \\
 \end{array}\right)\, ,
 \eeq
 and where $m_1 \gg m_2 \gg m_3$. Initially the model behaves according to the discussion of 
 Section~\ref{UVcompletion}; that is at the scale 
 \beq
\label{eq:lamir2}
\lir = \left(m_1 {\luv}^{\Delta_\pp-2 }\right)^{\frac{1}{\Delta_\pp-1}}\, ,
\eeq
we can integrate out the $\nf_1$ heavy quark 
 states. The theory can be dualized to a weakly coupled 
 and IR-free magnetic phase with $\nf_2+\nf_3$ flavours and a dynamical scale $\Lambda \sim \lir$. 
 However the magnetic superpotential is 
\beq
 \label{wmag2}
 \wmag= q \varphi \tilde{q} - \mu_\varphi^2 \varphi + \Lambda \varphi \Phi  - \mu_{\scriptstyle \Phi}^2 \Phi\, ,
 \eeq 
 where 
 \beq
\mu^2_\varphi=  \left( 
\begin{array}{ll} 
\mu_2^2 {\bf 1}_{\nf_2\times\nf_2} & {\bf 0}_{\nf_2\times \nf_3} \nn\\
 {\bf 0}_{\nf_3\times\nf_2} & \mu_3^2 {\bf 1}_{\nf_3\times\nf_3} 
  \end{array}\right)\, .
 \eeq
 The flavour contractions  are self-evident: for example the contraction $\varphi \Phi$ can involve only those 
 elements of $\varphi $ in the $\nf_2\times \nf_2$ upper block. As this is a mass term, the $\nf_2\times \nf_2$ blocks 
 of flavour adjoints may be integrated out supersymmetrically near the Landau pole scale $\Lambda$ to leave 
 a superpotential 
 \beq
 \label{wmag3}
 \wmag'= \frac{\mu_{\scriptstyle \Phi}^2}{  \Lambda}  q_2 \tilde{q}_2 +  q_3 \varphi_{33} \tilde{q}_3 + q_3 \varphi_{32} \tilde{q}_2+q_2 \varphi_{23} \tilde{q}_3- \mu_3^2 \varphi_{33}\,  ,
 \eeq 
where we now indicate the flavour blocks with indices. Finally, assuming that the mass for $q_2$ and $\tilde{q}_2$ dominates, i.e. that 
$\mu_\Phi^2/\Lambda \gg \mu_\varphi$, we may integrate out these fields as well, to find
 \beq
 \label{wmag4}
 \wmag''= q_3 \varphi_{33} \tilde{q}_3 - \mu_3^2 \varphi_{33}  \, .
 \eeq 
Since no gauge symmetry has been broken so far, this is an $\SU{\nf_2+\nf_3-\nc}$ O'Raighfeartaigh theory 
that has $\nf_3$ flavours of quarks. We may now take a large $\nc$ limit. In this case, remaining inside the correct ranges of flavours means that $\nf_1+\nf_2+\nf_3>\frac{3}{2}\nc$  and $ \nf_2+\nf_3 \geq \nc+1$ also become large. However there is no such constraint on either $\nf_3$ or $\nnc=\nf_2+\nf_3-\nc$ which may both be of order unity. Hence the IR theory can be weakly coupled. Note that this type of SUSY breaking could be inserted directly into the Higgs sector of the MSSM model in the previous subsection. 

\subsection{General gauge mediation}

The metastable SUSY breaking of the previous section lends itself to an RS1 implementation of the 
``simplified'' gauge mediation scenario discussed in Ref.~\cite{MN}. The result is a holographic version of 
general gauge mediation~\cite{Meade:2008wd}. To briefly recap, the 4D picture is as follows. Suppose that 
the supersymmetry breaking sector (i.e. the $\SU{\nf_3}$ sector above) contains no direct connection with 
the Standard Model gauge groups, $G_{SM}$, 
but that there is an additional pair of messenger fields $f,\tilde{f}$ that {\em are} charged under $G_{SM}$. 
The authors of Ref.~\cite{MN} argued that one can expect higher order operators to be generated in the 
underlying electric theory of the form 
\beq
\label{wmnelec}
\welec \supset \frac{(Q\tilde{Q})(f\tilde{f})}{M_X} + m_f f\tilde{f}\, ,
\eeq 
where $m_f$ is the messenger mass and 
$M_X$ is the scale of underlying physics, namely the mass scale of new modes in the 
theory that are exchanged between the messengers and the strongly coupled ISS sector. 
For convenience  we are now (and will henceforth) drop the $33$ indices that identify this as the $\SU{\nf_3}$ block.
In the low energy theory the SUSY breaking and mediation part of the superpotential becomes
\beq
\label{magterms}
\wmag \supset \frac{\Lambda}{M_X}{\pp\, f\tilde{f}} + m_f f\tilde{f} + q\pp\tilde{q} - \mu_\pp^2 \pp\, .
\eeq
As noted in Ref.~\cite{MN}, the first term is precisely the usual spurion interaction of ordinary gauge mediation. 
However the effective coupling $\frac{\Lambda}{M_X}$ can be very small since generally one expects $\Lambda\ll M_X$. 
The advantage of this suppression is that the $R$-symmetry breaking in the theory is under strict control. Of course 
the terms in Eq.\eqref{magterms} do explicitly break $R$-symmetry since $\pp$ has $R$-charge 2 (it appears with a linear term 
in the rest of the superpotential), however it is still approximately conserved because of the smallness of the coupling to the 
spurion.  An equivalent statement (as prescribed by Ref.\cite{Nelson:1993nf}) is that a new global SUSY preserving minimum is 
introduced but that it is so far away in field space that it could never disrupt the metastability of the SUSY breaking ISS sector. 
Indeed it is clear that the linear $\pp$ term can be set to zero if $\langle f \tilde{f}\rangle = -\mu_\pp^2 M_X/\Lambda $, 
but this can be much larger than the scale $\Lambda$, making it irrelevant to physics in the magnetic theory. The phenomenology of these models is similar to that of conventional gauge mediation (with the main difference being that 
the NLSP decay length is parametrically longer~\cite{Abel:2010vb}). 

The AdS equivalent of this type of mediation is as shown in Figure~\ref{simplified}. We begin with the 
ISS configuration of the previous section but add the elementary messenger fields on the UV brane.
 As we have seen an additional $\eta$ meson is required on the UV brane and this 
will in general mix with the $\varphi$ meson through its couplings to the bulk field $\Phi$. 
We will implicitly assume -- in order to justify having bulk gauge bosons -- that some of the fields in the 
non-supersymmetry breaking sector (i.e. the $\SU{\nf_1}\times\SU{\nf_2}$ sector above) also couple to the Standard Model gauge groups, but that only the messenger field couples to $\eta$. 
In addition we require $R$-symmetry to be broken in the UV theory which is represented by an explicit
mass term for the messengers. The brane superpotentials are then by comparison with Eqs.\eqref{wmnelec} and \eqref{magterms} given by 
\ba
\wuv & = & \eta f\tilde{f} + m_f f\tilde{f } + m_\eta \eta (\p-\p_0)  \nn \\
\wir  & = & { q} \pp\tilde{ q} - m_\pp { \pp}{ \p}\, .
\ea
As we saw this is the theory that remains after all of the confining physics described above has taken place, 
so that $q$,~$\tilde{q}$ and $\pp$ represent the fields in the low energy $\nf_3\times\nf_3$ flavour block.
Without the $\p_0$ term the anomaly-free $R$-symmetry of Table~\ref{sqcd0-1} is unbroken. 
As required by comparison to the ISS model, the $\p_0$ term induces an expectation value for $\p$ on the boundary
(equivalent to $m_Q$ in the underlying strongly coupled QCD theory by the bulk/boundary correspondence) 
because of the $|\partial\wuv/\partial \eta |^2$ term in the effective potential. (Note that, as we shall see in a moment, we cannot just set
$\partial\wuv/\partial \eta =0$ because supersymmetry is broken.) 
This leaves a residual but anomalous $R$-symmetry. The mass term for the gauginos then breaks the $R$-symmetry entirely on the UV boundary as in Ref.\cite{MN}, but the IR brane retains it. In this way the gravitational dual description makes it geometrically 
explicit that the approximate $R$-symmetry of the IR theory is an emergent phenomenon\footnote{See Ref.\cite{Abe:2007ki} for an alternative example of this phenomenon.}. 

Before presenting precise details, let us describe how we expect the suppressed mediation of Ref.\cite{MN} to 
operate in the gravitational dual description. As we have said, the boundary terms break supersymmetry and 
enforce a linear combination of the 4D $\eta$ and $\pp$ fields to zero. To see how this happens it is useful 
to temporarily disregard the effect of warping and consider the 4D theory whose superpotential is
simply the sum of $\wir$ and $\wuv$: 
\beq
W  =  \eta f\tilde{f} + m_f f\tilde{f }+ m_\eta \eta (\p-\p_0)  + { q} \pp\tilde{ q} - m_\pp { \pp}{ \p}\, .
\eeq
One can use the residual flavour symmetry to diagonalise the problem, and it is then easy to see that 
$\nnc$ diagonal components of all the $F$-terms can be set to zero by choosing $\p_{ii}=\p_{0} = \tilde{ q}_iq_i /m_\pp $ for $i=1\ldots \nnc$. The remaining $\nc$ contributions to the potential are  
\beq
V  \supset  \sum_{i=\nnc + 1}^\nf  m_\eta^2   (\p_{ii}-\p_0)^2  + m_\pp^2 { \p}_{ii}^2\, ,
\eeq
where to avoid confusion we are using the same symbol for the superfield and its scalar component.
Defining a mixing angle \beq \tan\vartheta = \frac{m_\pp}{m_\eta} \, , \eeq
the metastable ISS minimum occurs at
\beq 
\p_{ii}=\cos^2\vartheta  \,\p_0\,\, ; \, \, V_+(0)=\nc \p_0^2m_\eta^2 \sin^2 \vartheta \, ,
\eeq
with the effective $F$-terms for $\pp$ and $\eta$ being given by 
\ba
\mu_\pp^2 & = &  \sin\vartheta \cos\vartheta  \, m_\eta \p_0 \nn \\
\mu_\eta^2 & = &  \sin^2\vartheta \, m_\eta \p_0 \, ,
\ea
and with the remaining light meson field being given by 
\beq 
\pp' =\cos\vartheta \, \pp +  \sin\vartheta\,\eta  \, .
\eeq
The low energy theory is then 
\beq
\label{magterms2}
\wmag = \sin\vartheta\, {\pp'  f\tilde{f}} + m_f f\tilde{f} + q\pp'\tilde{q} + \mu_\pp^2 \pp'\, .
\eeq
Thus this purely perturbative 4D model is, at energies below $m_\eta$, essentially the configuration of \cite{MN} and the 
gauge mediation is standard. In particular note that the field $\p$ never contributes to supersymmetry breaking.

Now consider the strongly coupled theory modelled by a slice of AdS, depicted in Fig.\ref{simplified}. 
The supersymmetry breaking sector is similar to the model above, but altered in two ways. 
First, the field $\p$ is now a bulk field and generally 
has a profile that warps down the effective $\mu^2_\pp$. Second, the 
superpotentials themselves get an overall warp factor, which changes the relative sizes of the 
supersymmetry breaking contributions to the potential on the two branes. Nevertheless, 
some coarse aspects of the 4D model above carry over. For example the bulk meson $\p$ never contributes 
to the supersymmetry breaking, which is instead shared between the two branes.
In the limit of large $m_\eta$ the supersymmetry breaking will all be on the IR brane. 
Moreover the interesting 5D dynamics all happens above the scale $\lir$. Below this scale 
the theory behaves like the 4D one described above, modulo the warping of parameters. 
Therefore, if we choose a low messenger mass $m_f<\lir  $, then the mediation sector is oblivous to the 
strong coupling and the low energy phenomenology closely resembles that of Ref.\cite{MN}. The  
gaugino mass comes from the usual one-loop diagram and 
one finds the usual 4D result: 
\beq
\label{ggg}
M_\lambda = \sin\vartheta \frac{\alpha}{4\pi} \frac{\mbox{tr}(F_{\pp'})}{m_f }
\approx \sin\vartheta \frac{\alpha}{4\pi} \frac{\nc \mu_\pp^2}{m_f } \, ,
\eeq
in the limit of small $\vartheta$.
The scalar masses, being given by two loop diagrams, are similar in magnitude and, as in the 4D theory above, 
the phenomenology is similar to that of ordinary gauge mediation.

However new 5D effects {\em will} occur if we choose $m_f \gg \lir $. The scale $m_f$ then 
defines a resolution scale much smaller than the typical length scale corresponding 
to the Kaluza-Klein separation. The loop integrals that contribute to supersymmetry breaking 
are then effected by the localization of supersymmetry breaking on the IR brane. The net result is a  
suppression of the scalar masses with respect to the gaugino masses which are still given by Eq.\eqref{ggg}.
Naively one expects the suppression factor to be given by at least an extra loop factor for the scalars 
while the gauginos are from the AdS viewpoint a tree-level effect. This is nothing other than  
an AdS form of gaugino mediation very similar to that in Ref.\cite{Gherghetta:2000kr}.
It is remarkable that via the AdS/CFT correspondence, the simple model of Ref.\cite{MN} 
becomes a straightforward implementation of general gauge mediation~\cite{Meade:2008wd}!
(Note that the scalar mass-squareds in Ref.\cite{Gherghetta:2000kr} indeed 
conform to the general sum-rules derived in Ref.\cite{Meade:2008wd}.)
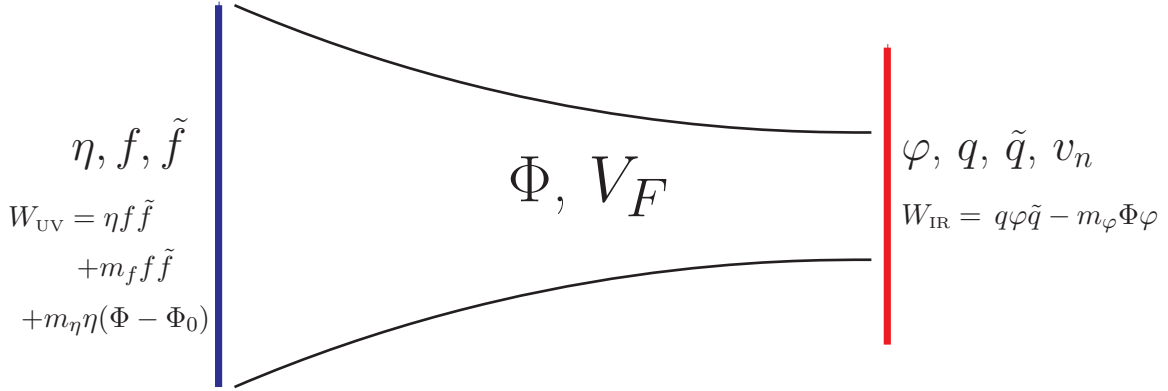
\begin{figure}[h]
\begin{center}
  \begin{picture}(420,148) (131,-190)
    \SetWidth{2.7}
    \SetColor{Blue}
    \Line(186,-44)(186,-188)
    \SetColor{Red}
    \Line(438,-60)(438,-172)
    \Text(296,-124)[lb]{\Huge{\Black{$\mbox{$\Phi$, $V_\nf$}$}}}
    \Text(445,-106)[lb]{\Large{\Black{$\mbox{$\varphi$, $q$, $\tilde{q}$, $v_\nnc$}$}}}
    \Text(131,-106)[lb]{\Large{\Black{$\mbox{$\eta,f,\tilde{f}$}$}}}
    \Text(107,-130)[lb]{{\small \Black{$\mbox{$\wuv=\eta f\tilde{f}$}$}}}
    \Text(133,-150)[lb]{{\small\Black{$\mbox{$+m_f f\tilde{f} $}$}}}
    \Text(112,-169)[lb]{{\small\Black{$\mbox{$+ m_\eta\eta (\p-\p_0) $}$}}}
    \Text(445,-130)[lb]{{\small\Black{$\mbox{$\wir=\, q\pp\tilde{q} -  m_\pp \p \pp$}$}}}
    \SetWidth{1.0}
    \SetColor{Black}
    \Bezier(192,-44)(304,-92)(400,-92)(432,-92)
    \Bezier(192,-188)(304,-140)(400,-140)(432,-140)
  \end{picture}
\end{center}
\caption{\label{simplified} \em The configuration for ``simplified'' gauge mediation (c.f. Ref~\cite{MN}).
Into the proposed gravitational dual of SQCD we add messenger fields $f,\tilde{f}$ on the UV brane that
interact with the bulk meson. The latter provides the heavy (KK) modes that generate the 
effective messenger/spurion coupling in the low energy theory.
\vspace{0.3cm}
}
\end{figure}

After this long heuristic discussion let us now present some precise details. The scales involved are $\lir \sim \zir^{-1}$ 
for the IR physics and $\luv\sim k=\zuv^{-1}$ for the UV physics, with the cuts-off being related as $\lir=\luv \zuv/\zir $.  
First the effect of the warping on the supersymmetry breaking: using the bulk solution for the massless 
modes given in Eq.\eqref{etavev}, and taking into account the warp factors in the canonical normalisation of $\varphi$, 
the low energy potential written in terms of $\puv=\p\zzuv$ is 
\beq
 V  \supset  \sum_{i=\nnc + 1}^\nf  m_\eta^2   (\puv-\p_0)^2  + \left(\frac{\zir}{\zuv }\right)^{-(1+2c)} m_\pp^2 \puv^2 \, ,
\eeq
and thus our mixing angle is 
\beq 
\tan\vartheta = \left(\frac{\zuv}{\zir }\right)^{\frac{1}{2}+c}  \frac{m_\pp}{m_\eta} 
\approx \left(\frac{\zuv}{\zir }\right)^{\frac{1}{2}+c} \, ,
\eeq
where the last relation is for the choice of $m_\pp\approx m_\eta$.
(The power of $(\frac{1}{2}+c)$ reflects the relation between the 
canonically normalised fields in Eq.\eqref{foobar}.)
Thus, when $c>-\frac{1}{2}$ $\tan\vartheta \ll 1$ and the supersymmetry breaking is pushed to the 
IR brane. Note that if on the other hand $c<-\frac{1}{2}$ 
then $\tan\vartheta \gg 1$. The supersymmetry breaking then naively looks to be completely localized on the UV brane, with $V_+= \nc \p_0^2 m_\eta^2 $, but in this case there is no metastability. Instead the non-perturbative terms discussed in Ref.\cite{ISS} introduce global supersymmetric minima 
at a distance less than $\lir$ away in field space, and so the Euclidean tunnelling  action is of order $S_E\sim 2\pi^2 \pp_{\mbox{\small min}}^4/V_+ \ll 1$: the supersymmetry breaking has to be an IR effect.

We can model the supersymmetry breaking gaugino masses with local $F$ terms on the branes. The effective operators are given by 
\beq 
W\supset  \frac{\auv}{g_5^2 \luv^2} \eta W^\alpha W_\alpha \delta (z-\zuv)
+\frac{\air}{g_5^2 \lir^2} \pp W^\alpha W_\alpha \delta (z-\zir)\, ,
\eeq
where $\auv, \air$ are constants.
To determine the value of the coefficients we need the 5D propagators which can be found in Refs.\cite{Gherghetta:2000qt,Gherghetta:2000kr,Marti:2001iw,Contino:2004vy}. The bulk gaugino propagator has a denominator of the form 
\beq
\tilde{J}_1\pzir\tilde{H}_1\pzuv - \tilde{H}_1\pzir\tilde{J}_1\pzuv\, ,
\eeq
where $H_1$ are Hankel functions of the first kind of order 1, the tilde modification is of the form 
\beq 
\tilde{J}_\alpha(w)=(-r + s/2 -1)J_\alpha(w) + w J_{\alpha -1 },
\eeq
where $s=1$ for gauginos, and where the values of $r$ for UV and IR branes respectively are  
\beq 
\ruv = -\frac{1}{2}+ {ip\zuv }\frac{\auv F_\eta }{4\luv^2}\,\, ; \,\, \rir = -\frac{1}{2}+  {ip\zuv }\frac{\air F_\pp}{ 4\lir^2}\, .
\eeq
It is straightforward to extract the pole of the propagator by taking the $p\zir , p \zuv \rightarrow 0$ limit, which gives a gaugino mass of 
\beq
\label{gauginomass}
 M_\lambda = \frac{\zuv^{-1}\luv^{-2}}{4\log{(\zir/\zuv)}} \left( {\air F_\pp }-{\auv F_\eta }\right)  \, .
 \eeq
As usual, the effect of the warping is to scale down the masses in the operator on the IR-brane by a 
factor $\lir/\luv$. Also note that no gaugino mass results if $\auv F_\eta$ is equal to $\air F_\varphi $: 
in the $F\rightarrow \infty $ limit this would correspond to having Dirichlet boundary conditions on both branes.  

For simplicity we shall henceforth neglect the small non-zero $F$-term that is induced on the UV brane, 
setting $\auv=0$ and focus on the IR-brane contribution.
By comparing with Eq.\eqref{ggg} and using the relation $g_5^2 k=g^2 \log(\zir/\zuv)$, we determine $\air$, 
to find that the gaugino mass is equivalent to an IR-brane localized term of the form 
\beq 
W\supset  
\sin\vartheta \frac{\zir^2}{\zuv^2} \frac{\pp}{4\pi^2 m_f} W^\alpha W_\alpha \delta (z-\zir)\, .
\eeq
Let us now compute the scalar mass-squared terms to see the suppression\footnote{Note that much of this discussion 
is valid for RS1 models with $F$-terms on the boundary in general.}. In order to do this we have to evaluate 
the one-loop contributions with bulk gauge fields in the loop. The mass-squared terms are given by 
\beq
m_i^2 = 4g^2 C(R_i) \Pi\, ,
\eeq
where $C(R_i)$ is the quadratic Casimir of the representation $R_i$ and 
\beq
\Pi = \frac{1}{k}\log (\zir/\zuv)\int \frac{{\rm d}^4p}{(2\pi)^4} \left[G_V(p,\zuv)-G_F(p,\zuv)\right] \, .
\eeq
The  gaugino, $G_F$ and gauge boson, $G_V$ propagators (which can be found in 
Ref.\cite{Gherghetta:2000kr}) are evaluated at $\zuv$, since the external squark fields are assumed to be localized
on the UV brane. For the gauge bosons we obtain the explicit expression 
\beq
G_V(p,\zuv) = \frac{1}{ip}\frac{J_1\pzuv Y_0\pzir- J_0\pzir Y_1\pzuv}{J_0\pzuv Y_0\pzir- J_0\pzir Y_0\pzuv}\, ,
\eeq
while the gaugino propagator is given by
{\small
\beq
G_F = \frac{1}{ip}\frac{J_1\pzuv Y_0\pzir- J_0\pzir Y_1\pzuv - \kappair(J_1\pzuv Y_1\pzir- J_1\pzir Y_1\pzuv)}
{J_0\pzuv Y_0\pzir- J_0\pzir Y_0\pzuv - \kappair(J_0\pzuv Y_1\pzir- J_1\pzir Y_0\pzuv)},
\eeq }where $\kappair = \frac{\air F_\pp}{ 4\lir^2} \frac{\zuv}{\zir}$ parametrizes the amount of supersymmetry breaking
on the IR brane.
Note that in the $\kappair\rightarrow \infty$ ($F_\pp\rightarrow \infty$) limit the gaugino wave-function is completely 
repelled from the IR-brane by the non-zero $F$-term, and one recovers the Green function for twisted boundary conditions, with Neumann (Dirichlet) boundary conditions on the UV (IR)-brane~\cite{Gherghetta:2000kr}, i.e.
\beq
G_F\, \stackrel{\mbox{\tiny$ F_\pp\rightarrow\infty$}}{=} \,\frac{1}{ip }\frac{{J}_1\pzir{Y}_1\pzuv- {J}_1\pzuv Y_1\pzir}{{J}_1\pzir{Y}_0\pzuv- {J}_0\pzuv{Y}_1\pzir}\, .
\eeq
The gaugino mass in this case is pure Dirac, and there are no divergences in $\Pi$. We refer to this as the 
``gaugino mediation limit''
\footnote{In order to calculate these well-known results using current correlators,
one would have to use the extended formalism of Ref.\cite{Benakli:2008pg} which includes 
Dirac masses. The purely Majorana piece gets exponentially suppressed as in~\cite{McGarrie:2010yk}.}.
However we are interested in the case when $F_\pp$ is finite.
In order to treat this more general case we use the following simplified
expression for the propagator difference: 
\beq 
   G_V(p,\zuv)-G_F(p,\zuv)= \frac{i}{p^3} \frac{\kappair}{\zir\zuv}\frac{1}{\Gamma_0}
   \frac{1}{(\Gamma_0+ i \kappair \Gamma_1)}\, ,
   \label{propdiff}
\eeq
where 
\ba
\Gamma_0 &=& I_0 \rpzuv K_0\rpzir -I_0\rpzir K_0\rpzuv \, , \nn \\
\Gamma_1 &=& I_0 \rpzuv K_1\rpzir +I_1\rpzir K_0\rpzuv \, .
\ea
The $I, K$ functions are modified Bessel functions and therefore $\Gamma_{0,1}$ are real valued. 
Using (\ref{propdiff}) we obtain (with $x=p \zir$).
\beq
i\Pi =  -\frac{\zir^{-2}}{8\pi ^2} \log{(\zir/\zuv)}
\int_0^\infty dx \, \frac{1}{\Gamma_0} \frac{i\kappair}{( \Gamma_0+ i \kappair\Gamma_1)}\, .
\label{pi0integral}
\eeq
The $i$ from the Wick rotation of $p_0 \rightarrow i p_0$ in $d^4p$ has been placed on the LHS of 
(\ref{pi0integral}). In the limit $\kappair\rightarrow\infty$ one obtains (using $\log (\zir/\zuv)= 34.54$)
\beq
i \Pi = -\frac{\zir^{-2}}{8\pi ^2} \log (\zir/\zuv)
\int_0^\infty dx\, \frac{1}{\Gamma_0 \Gamma_1}\approx (0.036)^2 \zir^{-2}.
\eeq
This corresponds to the real part of $i\Pi$ and reproduces the twisted boundary condition result in 
Ref.~\cite{Gherghetta:2000kr}.
Therefore the scalar mass-squared for finite $\kappair$ can be obtained by considering the real part
of $i \Pi$. Using (\ref{pi0integral}) we find
\beq
\label{piint2}
    \Re [i \Pi]= -\frac{\zir^{-2}}{8\pi ^2} \log{(\zir/\zuv)} \int_0^\infty dx \,\frac{\Gamma_1}{\Gamma_0} 
    \frac{\kappair^2 }{(\Gamma_0^2+ \kappair^2\Gamma_1^2)}\, .
\eeq 
In the limit $\kappair\rightarrow 0$ we find that $\Re [i \Pi] \propto \kappair^2$ as one would expect 
in normal gauge mediation. The ratio of the scalar masses to the gaugino masses can be parameterised by 
$\gamma $ such that  
\beq
    {\Pi} =\frac{ \gamma}{8\pi^2 }  {M_\lambda^2}  \, .
\eeq
(Numerically the twisted boundary condition result is equivalent to $\gamma = 1.73$.)
In the $\kappair\rightarrow 0$ limit we have 
\beq
       \gamma\simeq 
       -(\log(\zir/\zuv))^3  
       \int_0^\infty dx\, \frac{\Gamma_1}{\Gamma_0^3}\, .
       \label{gammaint}
\eeq
A part of this ratio comes from the RG running contribution of the Majorana gaugino masses to the 
scalar mass-squareds. Therefore, as one would expect, the integral (\ref{gammaint}) is logarithmically divergent
when $M_\lambda = 0$. In order to find the remaining piece we can compare $\gamma $ with the complete field theory expression for the contribution to the mass-squareds from each gauge factor (neglecting the 
running of the gauge couplings) \cite{Martin:1996zb}:
\beq
\label{martin}
\Pi_a(\mu) \approx \Pi_a(Q ) + \log\left(\frac{Q}{\mu}\right) \frac{M_{\lambda_a}^2}{8\pi^2} \, .
\eeq
The logarithmic piece in the integral for $\gamma $ exactly reproduces this RG running. Subtracting this piece, 
we find that in the large $\log{(\zir/\zuv)}$ limit the remaining finite contribution to $\Pi(Q)$ 
is given by
\beq 
\lim_{\mbox{\small
 ${z_{\mbox{\tiny IR}}/z_{\mbox{\tiny UV}}}\rightarrow\infty$}}\left[\bar{\gamma}\right]{=} \frac{1}{2} \log(\zir/\zuv)\, .
\eeq
Numerically, this approximation is accurate to a few percent for $\log{(\zir/\zuv)} = 34.54$ say. 
At first sight the apparent increase of $\bar{\gamma }$ with $\log{(\zir/\zuv)}$ is a bit puzzling since
heuristically one expects the supersymmetry mediation to scalars to tend to a constant, but actually this relation 
just reflects the ``messenger content'' in the bulk. Indeed this 
limit together with the AdS/CFT relation $g_5^2 k = 8\pi^2 /\bcft$ (c.f. Eq.\eqref{boundy}) gives
\beq
m_i^2 = \sum _a \frac{2 C_a}{\bcft} M_{\lambda_a}^2\, .
\eeq
This is the AdS/CFT equivalent to the perturbative extra-ordinary gauge mediation
relation where the contribution to the beta function coefficient is given by the 
number of messengers $N_{\mbox{\tiny mess}}$, and where we have  
$ M^2_\lambda \sim m_i^2 N_{\mbox{\tiny mess}}$~\cite{Cheung:2007es}. Here in the Veneziano limit the CFT contribution 
to the beta functions coefficients is instead $\bcft \sim \nc$. 

This result was in the $\kappair \rightarrow 0 $ limit, which we will refer to as the ``extra-ordinary gauge mediation limit''. 
However one can determine $\gamma $ and extract $\bar{\gamma}$ 
numerically for the general case as we increase $\kappair$. In order to do this we first note that when $\kappair \neq 0$ the integral is no longer divergent. This is to be expected since the gaugino is massive and 
therefore the logarithmic divergence has a natural IR cut-off about the gaugino mass scale. Indeed as $x\rightarrow 0$ 
the integrand in Eq.\eqref{piint2} can be approximated by 
\beq
\,\log (\zir/\zuv)\frac{\Gamma_1}{\Gamma_0} 
    \frac{\kappair^2 }{(\Gamma_0^2+ \kappair^2\Gamma_1^2)}
    \rightarrow 
    \frac{1}{x}\, 
\frac{\kappair^2}{\left(\log^2 (\zir/\zuv)+\frac{\kappair^2}{x^2}\right)}\, ,
\eeq
and this function is peaked at $x\equiv p\zir = \kappair /\log (\zir/\zuv)$, that is precisely where $p = M_\lambda$. The 
integrand and its approximation are shown in Figure~\ref{fig:poles}. 
The main feature of the ``extra-ordinary gauge mediation limit'' is that in this region 
the gaugino pole is well separated, so that 
one can define a ``messenger scale'', $Q_{\mbox{\tiny mess}}$, below which the contribution to the mass-squared integral is 
well described by the $\log(Q/\mu)$ piece in Eq.\eqref{martin}.
In order to find $\bar{\gamma}$ for arbitrary $\kappair$ therefore, we can divide the integral into two regions,
 with $\bar{\gamma}$ being identified with the contributions from above the scale $Q_{\mbox{\tiny mess}}$ where the two curves diverge. We can use the local minimum to define this point, whose location is well approximated by the value of the gaugino mass in the gaugino mediation (large $\kappair$) limit (i.e. it is at 0.24 TeV for the values chosen above).
This procedure ceases to be meaningful for large values of $\kappair$ because the pole ``melts'' into the main contribution. At this point extraordinary gauge mediation 
behaviour goes over to gaugino mediation behaviour as in Ref.\cite{Gherghetta:2000kr}. The numerically evaluated $\gamma $ and $\bar{\gamma}$ (in the extra-ordinary GM region) are shown in Fig.~\ref{fig:gamma}.
\begin{figure}[h]
\begin{centering}
\includegraphics[angle=0,scale=.58]{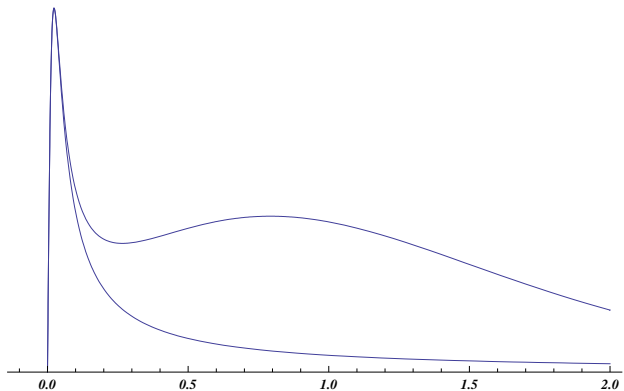}
\par\end{centering}
\caption{\it Contribution to the scalar mass-squareds with momentum in units of $\lir$: the full Green's function (upper curve) and the 
gaugino pole approximation (lower curve).} 
\label{fig:poles}
\end{figure}
\begin{figure}[h]
\begin{picture}(200,200) (-100,-20)
   \SetWidth{2.7}
\begin{centering}
 \Text(190,-5)[lb]{\mbox{$ \kappair $} }
 \Text(30,160)[lb]{\mbox{Extra-Ordinary GM} }
 \Text(200,30)[lb]{\mbox{Gaugino mediation} }
 \Text(125,117)[lb]{\Large\mbox{$\gamma $} }
 \Text(70,110)[lb]{\Large \mbox{$\bar{\gamma} $} }
\includegraphics[angle=0,scale=.35]{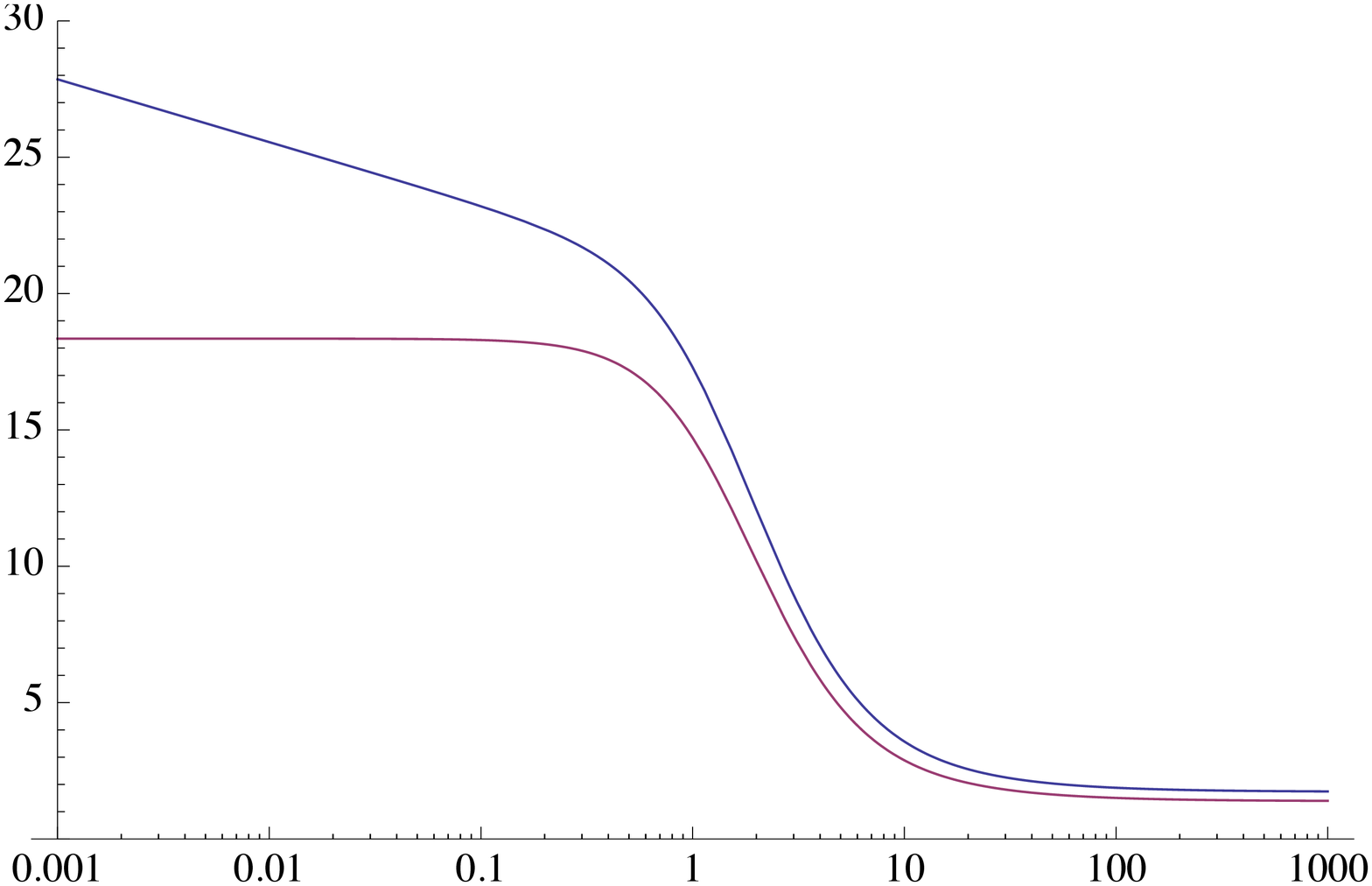}
\end{centering}
\end{picture}
\caption{\it The parameter $\gamma =   {8\pi^2 }{\Pi} /{M_\lambda^2} $,
varying continuously from extraordinary gauge mediation to gaugino mediation behaviour as 
 the relative supersymmetry breaking on the IR brane, $\kappair $, increases. The $\bar{\gamma}$ line, representing the mass-squared value ``at 
the messenger scale'',  is the contribution with the gaugino RG term removed, relevant in the 
small $\kappair $ limit.} 
\label{fig:gamma}
\end{figure}

As one final remark, it is worth highlighting the restricted form of general gauge mediation
that one derives from this model. It is by now well known that the most general configuration for gauge 
mediation allows six independent parameters (assuming no CP violating phases in the gaugino sector), 
three for the gaugino mass terms, and three for the squarks~\cite{Meade:2008wd}. 
There are five squark masses in total so this requires two sum rules,
\ba 
m^2_{\tilde Q} -2m^2_{\tilde U} +m^2_{\tilde D} -m^2_{\tilde L} +m^2_{\tilde E} &=& 0 \nn\\
2m^2_{\tilde Q} -m^2_{\tilde U} -m^2_{\tilde D} -2m^2_{\tilde L} +m^2_{\tilde E} &=& 0\, .
\ea
The squark masses derived here and in Ref.\cite{Gherghetta:2000kr} (which are realisations of general gauge mediation in AdS) of course have to satisfy these rules. However there are only four 
free parameters for the models discussed here, not six. Assuming that the gaugino masses are driven by couplings to different $F$-terms or 
possibly different couplings to the same $F$-term, then they can be free parameters, however the mediation 
to the sfermions is only a function of the AdS geometry and the suppression is the same for all the Standard Model 
gauge factors. Therefore the pattern of soft-supersymmetry breaking can be written in terms of the gluino mass $M_3$ and three sfermion mass-squared parameters $\Pi_{i=1\ldots 3}$ as follows:
\ba
M_1 = \sqrt{\frac{\Pi_1}{\Pi_2}} M_2 &=& \sqrt{\frac{\Pi_1}{\Pi_3}} M_3 \nn \\
m^2_{\tilde Q} & = & \frac{4}{3} \alpha_3 \Pi_3 +\frac{3}{4} \alpha_2 \Pi_2+\frac{1}{60} \alpha_1 \Pi_1  \nn \\
m^2_{\tilde U} & = & \frac{4}{3} \alpha_3 \Pi_3 +\frac{4}{15} \alpha_1 \Pi_1  \nn \\
m^2_{\tilde D} & = & \frac{4}{3} \alpha_3 \Pi_3 +\frac{1}{15} \alpha_1 \Pi_1  \nn \\
m^2_{\tilde L} & = & \frac{3}{4} \alpha_2 \Pi_2 +\frac{3}{20} \alpha_1 \Pi_1  \nn \\
m^2_{\tilde E} & = & \frac{3}{5} \alpha_1 \Pi_1 \, . 
\ea

\section{Conclusions}

We have examined Randall-Sundrum (RS1) like configurations in strongly coupled 4D ${\cal N}=1$ 
supersymmetric field theory. By taking a large $\nc$ (Veneziano) limit and combining it with a Seiberg 
duality, we showed how one can construct a model in which a conformal phase with relevant operators 
(specifically quark mass terms) flows to a weakly coupled free-magnetic phase. The bulk of these theories is approximated by the construction of Klebanov and Maldacena~\cite{Klebanov:2004ya}. The magnetic theory, including its gauge fields, lives entirely on the IR brane as emergent degrees of freedom.

We showed how this construction can be used to derive an RS1 version of the MSSM in which 
the $\SU{2}_L$ gauge group is emergent. The $\SU{3}_c$ and hypercharge gauge bosons are 
bulk degrees of freedom and correspond to part of the ``flavour'' symmetries of the Seiberg duality. 
The  right-handed fields are predicted to be entirely elementary, whereas the left-handed fields are 
predicted to be a mixture of elementary and composite degrees of freedom. (The latter are identified as
the mesons of the Seiberg duality.)

We also showed how gaugino mediation can be implemented, by beginning with the Murayama-Nomura model 
of gauge mediation in Ref.~\cite{MN} and taking its large $\nc$ limit in the specified manner. The metastable 
supersymmetry breaking of Ref.~\cite{ISS}, being an emergent phenomenon, appears on the IR brane, while the 
matter fields and messenger fields (being elementary degrees of freedom in the model) are on the UV brane. The 
Standard Model gauge fields are bulk degrees of freedom and therefore gauginos get masses at leading order, 
whereas the sfermion mass-squareds,  which have to be transmitted through the bulk, are suppressed. 
The result is an AdS version of extra-dimensional gauge mediation. 
By varying parameters, the pattern of supersymmetry breaking can be taken from extra-ordinary gauge mediation 
(i.e. equivalent to a large number of messengers that are integrated out below a typical mass scale) to AdS 
gaugino mediation similar to that of Ref.\cite{Gherghetta:2000kr}. 
Due to the universal nature of the mediation, the model corresponds to  
general gauge mediation (with additional Dirac gaugino masses) but with only four free parameters. 

\subsection*{Acknowledgements} We thank  Mark Goodsell and Jose Santiago for discussions. 
SAA gratefully acknowledges the Royal Society and the University of Melbourne, and 
TG the IPPP Durham, for support and hospitality. We also thank Yann Mambrini and Orsay, where 
part of this work was done. SAA is supported by a Leverhulme Senior Research Fellowship and 
the European RTN grant ``Unification in the LHC era'' (PITN-GA-2009-237920). TG is supported 
by the Australian Research Council.

\end{document}